\documentclass[
    reprint, 
    aps,
    preprintnumbers,
    twocolumn,
    prb,
    superscriptaddress
]{revtex4-2}

\usepackage[utf8]{inputenc}
\usepackage{booktabs}
\usepackage[multiple]{footmisc}
\usepackage{lipsum}
\usepackage{rotating}
\usepackage{lipsum}

\usepackage{amssymb}
\usepackage{amsbsy}
\usepackage{amsmath}
\usepackage{bm}
\usepackage{bbm}
\usepackage{xfrac}
\usepackage{mathtools}

\usepackage{tikz}
\usepackage[T1]{fontenc}
\usepackage{etoolbox}
\usepackage{graphics}
\usepackage{multirow}
\usepackage{url}
\usepackage{hyperref}
\usepackage{placeins}
\usepackage[normalem]{ulem}

\usepackage{color}

\usepackage{adjustbox}

\graphicspath{{mainplots/}} 

\newcommand*\vc[1]{\mathbf{#1}}
\newcommand*\tx[1]{\mathrm{#1}}
\newcommand*\wn{cm$^{-1}$}

\newcommand{\1}[1]{#1}

\begin{document}

\title[]{From Enhanced Sampling to Human-Readable Representations of Protein Dynamics}

\author{Souvik Mondal}
\thanks{Current address: Department of Basic Science \& Humanities, Institute of Engineering \& Management, School of UEM, Kolkata 700091, West Bengal, India}
\author{Michael A. Sauer}
\author{Matthias Heyden}
\email{mheyden1@asu.edu}
\affiliation{School of Molecular Sciences, Arizona State University, Tempe, AZ 85287, U.S.A.}

\date{\today}

\begin{abstract}
Understanding protein conformational dynamics is essential for elucidating biological function but remains challenging due to the wide range of timescales and the complexity of collective motions. Enhanced sampling methods overcome timescale limitations of conventional molecular dynamics, yet their effectiveness depends on the choice of collective variables (CVs), which are often difficult to define and may lack physical interpretability. In particular, collective variables derived from machine learning or collective vibrational modes can efficiently capture slow dynamics but are not easily mapped onto intuitive structural descriptors.
Here, we present a fully automated framework that transforms enhanced sampling trajectories into human-readable representations of protein dynamics. Our approach combines enhanced sampling along CVs derived from frequency-selective anharmonic mode analysis with a post hoc analysis of biased trajectories using weighted dynamic cross-correlation matrices. From these, we identify residue pairs and domains exhibiting correlated and anti-correlated motions, yielding simple domain-domain distances that serve as physically interpretable CVs.
We apply this method to five proteins, including KRAS and HIV-1 protease, and show that it consistently identifies biologically relevant domains and motions without prior system-specific knowledge. Projection onto these distances produces free energy surfaces that reproduce known conformational states with low statistical uncertainty while maximizing independent dynamical information.
This workflow enables systematic recasting of complex CVs into simple geometric descriptors without loss of essential dynamics. Its generality and automation make it broadly applicable for interpreting enhanced sampling simulations and generating interpretable conformational ensembles for integration with emerging machine learning approaches.
\end{abstract}

\maketitle

\section{Introduction}
\label{s:intro}

Protein dynamics is essential for many biological functions and involves a wide range of timescales -- from fast bond vibrations in femtoseconds to slow conformational changes over milliseconds\cite{henzler2007,wedemeyer2002,boehr2006}. 
Identifying key residues or domains that drive these dynamics is vital to understanding functional mechanisms and offers significant potential for the rational design of enzymes with custom activity or the development of highly specific allosteric drugs.

Molecular dynamics (MD) simulations are a powerful approach to capture atomistic details of protein motion\cite{karplus2002}. 
Although advances in computational techniques and hardware have extended accessible simulation timescales, conventional MD simulations still face significant challenges in exploring rare or slow conformational transitions, especially those occurring beyond the microsecond regime\cite{dror2012,shaw2021,ayaz2023,greisman2023}.
Novel machine learning models can help to overcome such limitations but rely on the availability of sufficient training data.\cite{bioemu}
To overcome the timescale limitations of unbiased MD, many enhanced sampling techniques have been developed over the years\cite{bernardi2015,mitsutake2001,torrie1977,isralewitz2001,schlitter1994,henin2004,laio2002,sugita1999,hamelberg2004}.
Most of these techniques rely on collective variables (CVs) to efficiently drive transitions of interest. 
The selection of these CVs usually relies on physical intuition or experimental observations\cite{mehdi2024enhanced}. 
However, when the relevant conformational changes are unknown or complex, defining suitable CVs can become a major challenge. 
As a result, the performance of CV-based methods strongly depends on the quality of the chosen variables -- poor choices can lead to inefficient sampling and incomplete exploration of the energy landscape\cite{mehdi2024enhanced}.

Various methods are used to identify suitable collective variables (CVs) based on structural or dynamic data from unbiased simulations. 
Structure-based approaches focus on low-frequency vibrations that are often associated with large-amplitude motions of biomolecules. 
These methods typically apply harmonic normal mode analysis to a model of the system's potential energy, either coarse-grained (e.g., elastic network models) or atomistic\cite{go1983,aalten1995}. 
However, for the lowest frequencies of a biomolecular system, the validity of harmonic approximations is intrinsically limited.\cite{hayward1995,ma2005,sauer2023frequency}
Alternatively, dynamics-based approaches such as Principal Component Analysis (PCA)\cite{go1983,garcia1992large,amadei1993essential} and Time-lagged Independent Component Analysis (TICA)\cite{naritomi2013,noe2015kinetic} extract collective variables (CVs) from simulation data, potentially avoiding the limitations of harmonic approximations. 
PCA identifies CVs that correspond to the largest variance, representing the most significant motions observed in the simulation\cite{go1983,garcia1992large,amadei1993essential}, while TICA focuses on extracting CVs associated with the slowest dynamics\cite{naritomi2013,noe2015kinetic}. 
Despite their advantages, these methods often encounter challenges related to sampling limitations in the available simulation trajectories and the derived CVs may not be consistently reproducible in independent simulations.

Machine learning (ML) methods provide novel powerful tools to identify collective variables (CVs) from simulation data\cite{sidky2020machine,wang2020machine,noe2020machine,hanni2025data}. 
They excel at extracting low-dimensional representations of slow dynamics from extended simulations. 
A key advantage of ML-derived CVs is their ability to capture nonlinear combinations of geometric variables, often outperforming manual approaches.\cite{fu2024collective}
Further, deep learning can be applied directly to simulation trajectories without the need for custom feature engineering.\cite{frohlking2024deep} 
However, many ML methods still rely on extensive simulation data for training. 

We recently developed a high-throughput method for enhanced conformational sampling that requires only short simulations on the nanosecond timescale to reliably identify effective CVs.\cite{mondal2024exploring,sauer2025high,sauer2026fast}
Our approach is based on a FREquency-SElective ANharmonic (FRESEAN) mode analysis of molecular vibrations that isolates low-frequency vibrations and collective degrees of freedom associated with conformational change.\cite{sauer2023frequency}
We identify CVs suitable for conformational sampling through zero frequency fluctuations that describe diffusive behavior and solvent damping due to modulated protein-solvent interactions.\cite{neff26protein}
Metadynamics simulations based on our CVs generate free energy surfaces with small statistical errors that reproduce the known behavior for a diverse set of enzymes.\cite{sauer2026fast}.
For direct comparisons to data from the literature, we unbias metadynamics trajectories and project them into geometric variable spaces (distance, angles, etc.) used in previous literature.
After unbiasing and projecting our simulations, we obtain free energy surfaces with statistical errors on the order of $k_B T$ that demonstrate the effectiveness of our approach to robustly enhance conformational sampling.\cite{sauer2026fast}

However, without a known space of geometrical variables, {\em e.g.}, to compare to experiments or prior simulations, interpreting free energy surfaces in a space described by complicated CVs can be challenging.
This is equally true for non-linear CVs generated with ML methods or CVs describing the concerted motions of many atoms such as low-frequency vibrations.
In such cases, it is desirable to recast the free energy surface into a set of simpler 'human-readable' variables that directly represent geometrical measures, {\em e.g.}, distances.
Here, we propose a straightforward approach to automatically identify residue-residue or domain-domain distances related to the conformational changes observed in biased trajectories generated in enhanced sampling simulations with an independent set of CVs.
Our approach utilizes weighted averages over biased trajectories to compute a dynamic cross correlation matrix (DCCM)\cite{mccammon1984protein} from which we extract residues and domains that move collectively or against each other using a simple algorithm.
Our application to a set of previously studied test systems shows that this approach, combined with enhanced sampling along CVs obtained from FRESEAN mode analysis, allows for a 'human-readable' characterization of protein collective dynamics without the need of any prior system-specific knowledge.

\section{Theory}
\label{s:theory}

\subsection{Dynamic Cross-Correlation Matrix from Biased Trajectories}

To identify correlated motions in our biased simulations, we select $C_\alpha$ atoms from each amino acid residue and remove translational and rotational motion from the trajectory by minimizing the root mean squared deviation (RMSD) relative to a reference structure.
We then compute the average protein structure $\langle \vc{r} \rangle$ of the unbiased ensemble as a weighted expectation value over the time steps $t_s$ of the biased simulation trajectory.
\begin{equation}
\langle \vc{r} \rangle = \frac{\sum_s w(t_s) \, \vc{r}(t_s)} {\sum_s w(t_s)}
\end{equation}
The weights $w(t_s)$ correct for the bias potential applied at coordinates $r(t_s)$ (defined after projection into low-dimensional CV-space).
\begin{equation}
    w(t_s) = e^{V_\tx{bias}\left[\vc{r}(t_s)\right]}
    \label{e:w}
\end{equation}

We then define the displacements from the average structure as $\vc{\Delta r}_t = \vc{r}_t - \langle \vc{r} \rangle$, which can be split into three-dimensional displacement vectors $\vc{\Delta r}_{t,i}$ for the $C_\alpha$ atom of each residue $i$.
The dynamic cross-correlation matrix (DCCM) is then defined as:
\begin{equation}
\label{e:corr}
C_{ij}=\frac{\langle \Delta r_i \cdot \Delta r_j \rangle}{\langle \Delta r_i \cdot \Delta r_i \rangle^{\frac{1}{2}} \cdot \langle \Delta r_j \cdot \Delta r_j \rangle^{\frac{1}{2}}} \,\,,
\end{equation}
where each average is again computed as a weighted expectation value:
\begin{equation}
\langle \Delta r_i \cdot \Delta r_j \rangle = \frac{\sum_s w(t_s) \, \Delta r_i(t_s) \cdot \Delta r_j(t_s)} {\sum_s w(t_s)}
\end{equation}
The resulting entries $C_{ij}$ in the DCCM range from $-1$ to $+1$ with positive (negative) values that indicate correlated (anti-correlated) motion of residue pairs.  

For the analysis of multiple independent metadynamics simulations, we constructed the DCCM from overall averages of $\langle \vc{r} \rangle$ and $\langle \Delta r_i \cdot \Delta r_j \rangle$ over all trajectories. 
 
\subsection{Identifying Correlated and Anti-Correlated Domains}

To identify residue and domain pairs that best represent the collective dynamics observed in our enhanced sampling simulations, we first calculated separate sums of positive and negative elements for each row $i$ of the symmetric DCCM.
\begin{eqnarray}
    z_i^+ &=& \sum_j \left( C_{ij} | C_{ij} > 0\right) \\
    z_i^- &=& \sum_j \left( C_{ij} | C_{ij} < 0\right)
\end{eqnarray}
Rows with large sums of positive elements correspond to residues in collectively moving domains. 
Rows with large (negative) sums of negative elements correspond to residues that move against a collectively moving domain. 
We aim to identify residues and residue pairs that have both properties.
Our first selected residue corresponds to the index $i$ that maximizes $z_i^+ - z_i^-$ and we label it $a_1$.
Based on our selection, residue $a_1$ belongs to a collectively moving domain that moves against another collectively moving domain. 
We then simply select the residue that is most anti-correlated to residue $a_1$ by identifying the index $a_2$ of the most negative element in row $a_1$ of the DCCM.
This selection yields a pair of residues that are each part of collectively moving domains that exhibit anti-correlated dynamics.
The distance between these residues provides a suitable collective variable to characterize this collective motion with a simple variable.
To further reduce the impact of local fluctuations, we can identify the other residues that are part of the collectively moving domains by analyzing rows $a_1$ and $a_2$ of the DCCM. 
Using a simple threshold value (we use $+0.5$), we select the residues whose dynamics are most correlated with residues $a_1$ and $a_2$, respectively. 
The residues form the domains $A_1$ and $A_2$ and we use the distance between their centers of mass to describe their collective anti-correlated motion.

Next, we aim to identify residues that exhibit dynamics that is maximally uncorrelated to the residue pair $a_1$ and $a_2$ to generate a second independent variable. 
For this purpose, we remove the sign of the values in rows $a_1$ and $a_2$ of the DCCM and add them. 
In the resulting list, we identify the minimum value (least correlations with $a_1$ and $a_2$) and label the corresponding index $b_1$. 
As before, we identify the residue that is most anti-correlated to residue $b_1$ by identifying the index $b_2$ of the most negative element in row $b_1$ of the DCCM.
This selection of $b_1$ and $b_2$ prioritizes the absence of correlations to residues $a_1$ and $a_2$.
Alternative choices are possible that place a greater emphasis on residues $b_1$ and $b_2$ being part of a collectively moving domain. 
However, despite not prioritizing collective motion in the selection of residues $b_1$ and $b_2$, we frequently observed that both are part of collectively moving domains using the same formalism described above (for domains $A_1$ and $A_2$) to define the domains $B_1$ and $B_2$.

Thus, independent of the complexity of the CVs used to enhance sampling, 
we can now characterize the sampled conformational changes using simple domain-domain distances $\left(A_1-A_2\right)$ and $\left(B_1-B_2\right)$.

\subsection{Free Energy Surfaces in Distinct Variable Spaces}
Recasting free energy surfaces from one set of variables (the CVs used during enhanced sampling) into another (here: domain-domain distances) is straightforward. 
The weights of each simulation time frame defined in Eq.~\ref{e:w} can be combined with the evaluations of the new set of CVs, {\em e.g.}, $d_A = \tx{distance}\left(A_1-A_2\right)$ and $d_B = \tx{distance}\left(B_1-B_2\right)$ to construct a weighted histogram in the corresponding space.
After normalization, this results in the unbiased probability distribution as a function of the new variables:
\begin{equation}
    p\left(d_A,d_B\right) = \frac{\sum_s w(t_s) \, \delta\left(d_A(t_s)-d_A\right) \, \delta\left(d_B(t_s)-d_B\right)} {\sum_s w(t_s)} \,\,,
    \label{e:prob}
\end{equation}
where the Kronecker $\delta$ function is used to assign time frames to a specific bin of the histogram.
 Normalized histograms or probability distributions can be averaged over multiple simulation trajectories. 
The result can then be converted back into a free energy surface defined by the new set of CVs:
\begin{equation}
    \Delta G\left(d_A,d_B\right) = - k_B T \, \tx{ln}\left[p\left(d_A,d_B\right)\right]
    \label{e:fes}
\end{equation}

\section{Results and Discussion}
\label{s:results}
Our test systems are five proteins that we have studied previously using enhanced sampling simulations biased along anharmonic low-frequency vibrations.\cite{sauer2026fast}
These proteins are hen egg-white lysozyme (HEWL), HIV-1 Protease (HIV-1 Pr), myeloid cell leukemia 1 (MCL-1), ribose binding protein (RBP), and Kirsten rat sarcoma virus (KRAS).
Compared to our previous work, where we used geometric variables from the literature to recast free energies into a 'human-readable' format, here we propose an automated strategy to extract such variables directly from the biased simulations without the need of prior knowledge. 
For brevity, we focus our following analysis on a single system, KRAS, in the main text and provide the corresponding results for the four other proteins in the \1{Supporting Information (SI)}.

\subsection{Dynamic Cross Correlations}
\begin{figure}[ht!]
\begin{center}
\includegraphics[clip=true,width=0.8\linewidth]{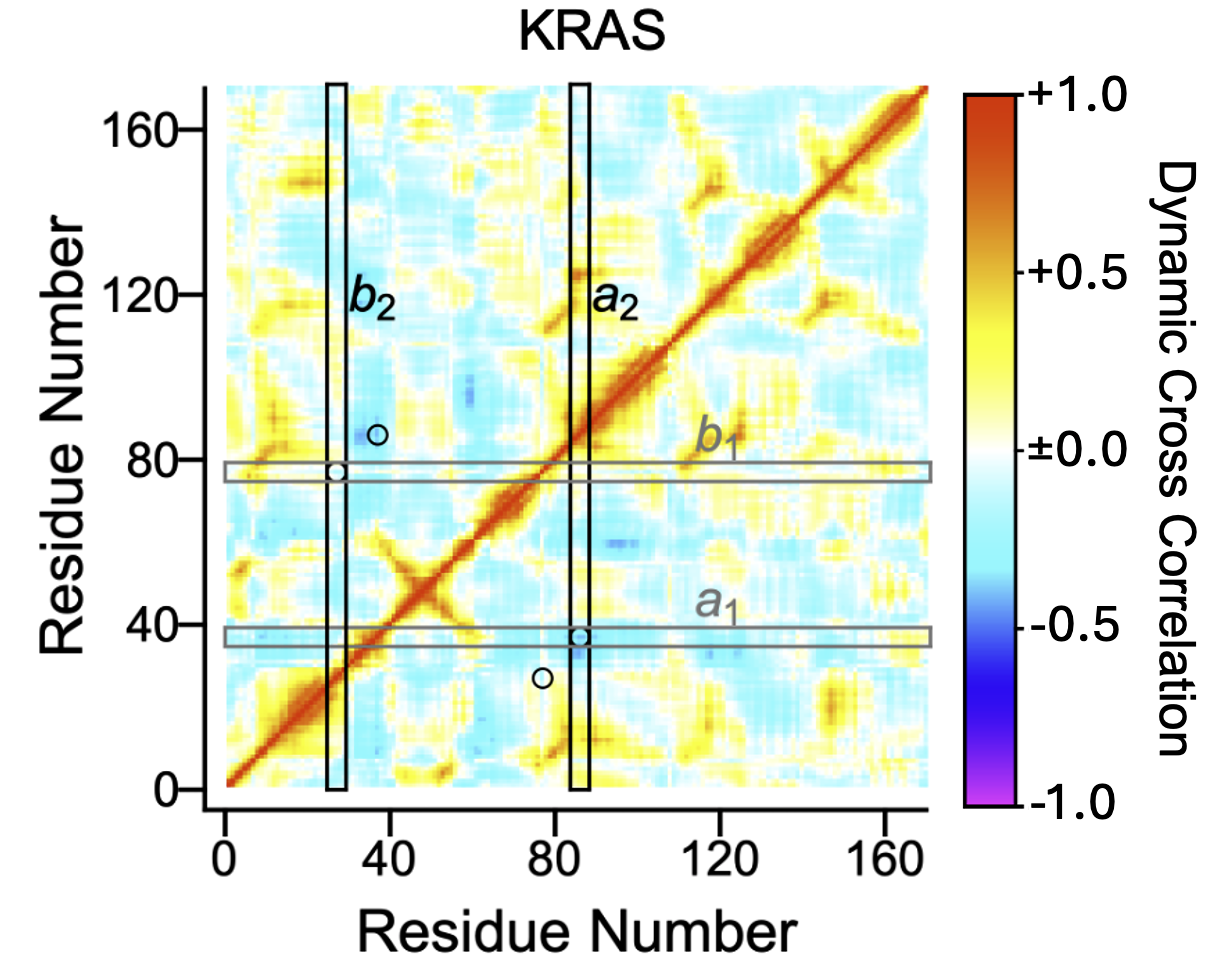}
\caption{
DCCM computed as a weighted average from enhanced sampling simulations of the protein KRAS.
We highlighted the selected anti-correlated residue pairs $a_1/a_2$ and $b_1/b_2$ with circles and the corresponding rows ($a_1$ and $b_1$) and columns ($a_2$ and $b_2$) with light and dark gray rectangles, respectively.
\1{See Figure~S1 in the SI for DCCMs obtained for HEWL, HIV-1 Pr, MCL-1 and RBP.}
}
\label{f:dccm}
\end{center}
\end{figure}

In Figure~\ref{f:dccm}, we plotted the average DCCM obtained from 20 trajectories generated in separate 100~ns well-tempered metadynamics simulations of KRAS.
In these simulations, we used low-frequency vibrations as CVs to define the bias potential and computed the DCCM from weighted averages as described in the Methods section. 
Additional simulation details are given in Ref.~\citenum{sauer2026fast} and analogous results for HEWL, HIV-Pr, MCL-1 and RBP are shown in \1{Figure~S1 of the SI}.

Using the algorithm described in the Theory section, we identified the residue pairs $a_1=$Ile36 and $a_2=$Asn85 as well as $b_1=$Glu76 and $b_2=$Asn26, which participate in collective and anti-correlated dynamics. 
The matrix elements corresponding to both residue pairs are indicated with circles in Figure~\ref{f:dccm}.
Notably, our algorithm does not simply select the most anti-correlated residue pair, which would correspond to the most negative element in the DCCM. 
Instead, the $a_1/a_2$ pair corresponds to the most collective anti-correlated motion in the simulated protein as expected for protein domains moving against each other.
We select the anti-correlated $b_1/b_2$ pair based on the lack of correlation with the $a_1/a_2$ pair.

\begin{figure}[ht!]
\begin{center}
\includegraphics[clip=true,width=1.0\linewidth]{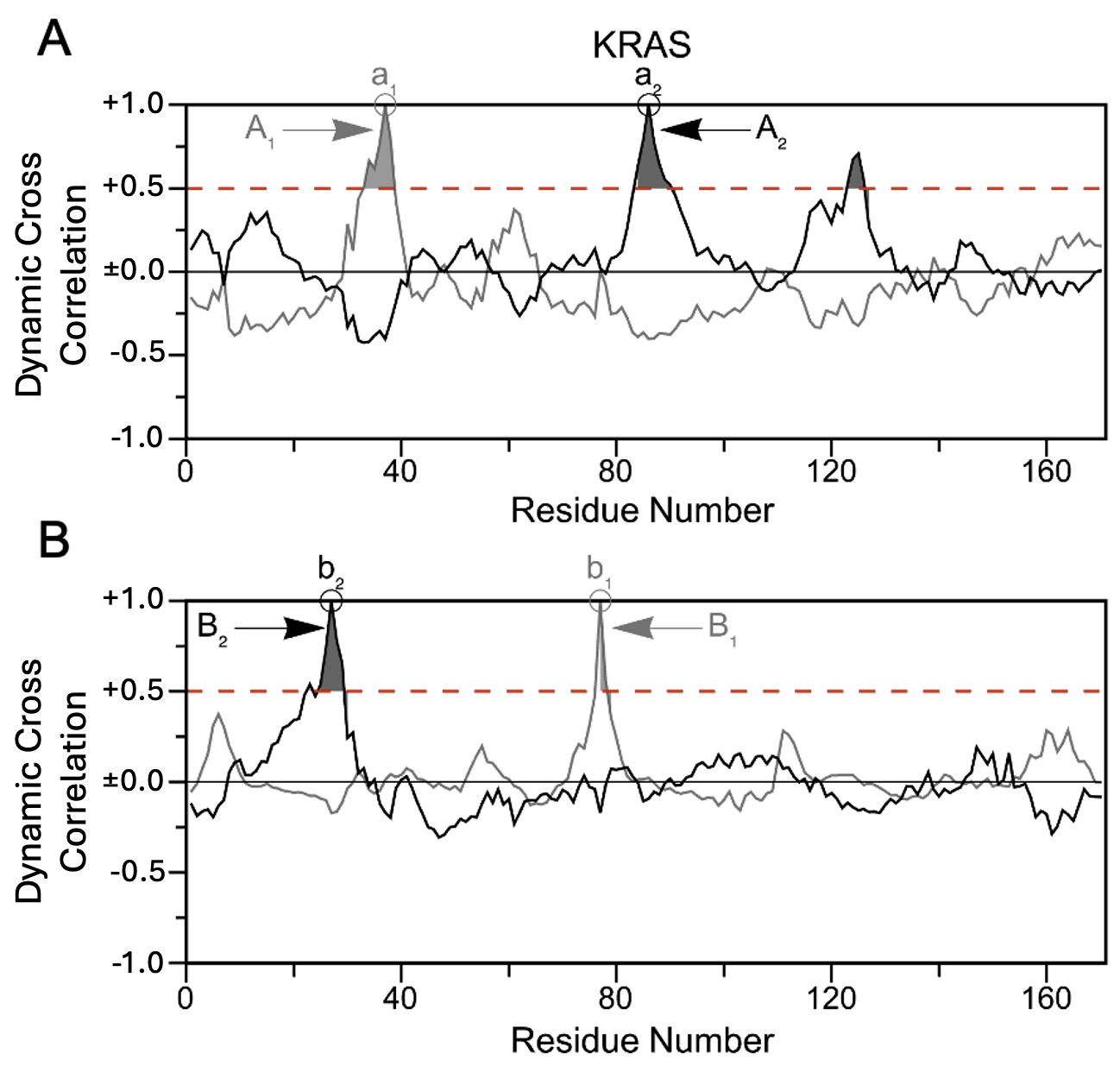}
\caption{
Numerical values in rows/columns (A) $a_1=36$ and $a_2=85$ and (B) $b_1=76$ and $b_2=26$ of the DCCM, each in light and dark gray, respectively
While self-correlations are 1 by definition, numerical values for other amino acids quantify their correlations with residues (A) $a_1$ and $a_2$ (A) and (B) $b_1$ and $b_2$. 
We use a simple threshold criterion indicated by the dashed red line to identify residues that form collectively moving domains $A_1$, $A_2$, $B_1$ and $B_2$ with residues $a_1$, $a_2$, $b_1$ and $b_2$ in their center.
The selected residues for each domain are highlighted by a shaded area under the corresponding plot for correlations that exceed the threshold.
Both plots also quantify the anti-correlation between the $a_{1/2}$ and $b_{1/2}$
\1{See Figure~S2 in the SI for DCCM columns and rows obtained for HEWL, HIV-1 Pr, MCL-1 and RBP.}
}
\label{f:corr}
\end{center}
\end{figure}

The numerical values in the rows/columns of the symmetric DCCM for selected residues $a_1$, $a_2$, $b_1$ and $b_2$ (highlighted in Figure~\ref{f:dccm}) allow us to easily identify residues that participate in collective motion. 
We plotted these numerical values in Figure~\ref{f:corr}.
Using a simple threshold value of +0.5, we identified residues forming the collectively moving domains $A_1$, $A_2$, $B_1$ and $B_2$ associated with each of the selected residues.
Notably, the residues of collectively moving domains are not necessarily close in sequence space. 
For example, domain $A_2$ for KRAS consists of two sets of amino acids (residues 83-89 and residues 122-125) that are separated in sequence but in direct contact with each other in the folded structure.
We identified at least one collectively moving domain formed by non-sequential sets of amino acids for each of the other four proteins as well (\1{Figure~S2 in the SI}).

\subsection{Collective Domains}

We verified in our previous work that the enhanced sampling simulations for each system reproduce known conformational transitions\cite{sauer2026fast}. 
Thus, the collectively moving domains identified by the algorithm presented here should provide alternative descriptions of these known transitions.
For example, we correctly identify the moving parts of the $\alpha-$ and $\beta-$domains involved in the lid-opening of HEWL as domains $A_1$ and $A_2$.
Similarly, we identify residues corresponding to the flap, flap elbow, and cantilever domains in each HIV-1 Pr monomer as $A_1$ and $A_2$, which move in opposite directions during the close-to-open transition of the enzyme\cite{hornak2006hiv}.

\begin{figure}[ht!]
\begin{center}
\includegraphics[clip=true,width=1.0\linewidth]{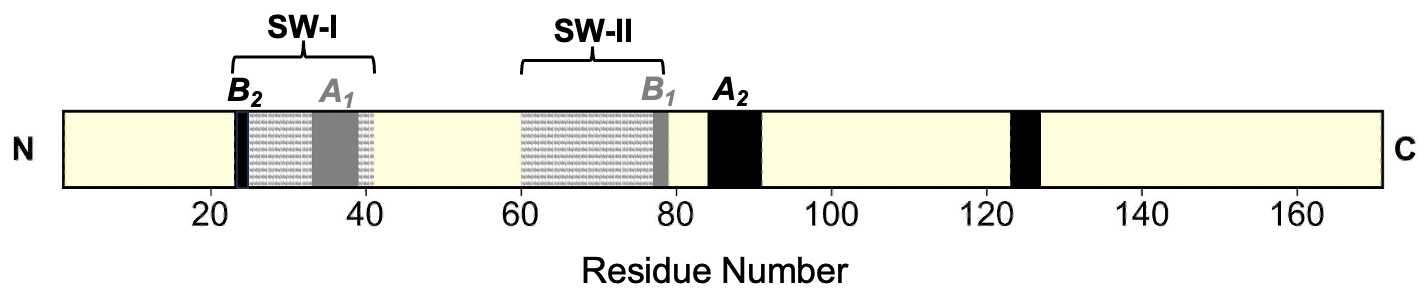}
\caption{
Location within the KRAS sequence of domains $A_1$, $A_2$, $B_1$ and $B_2$ as identified by their collective dynamics using our algorithm described in the Theory section. 
}
\label{f:seq}
\end{center}
\end{figure}

For KRAS, our analysis readily identified residues in the highly conserved switch I and switch II domains, which are essential for its function and known to be dynamic.\cite{pantsar2020current} 
As shown in Figure~\ref{f:seq}, we identified residues 32-37 as $A_1$ that overlap with the switch I domain of KRAS (residues 25-40).
Notably, more stringent definitions of the switch I domain of KRAS in the literature point only to residues 30-40~\cite{pantsar2020current} or residues 30-38~\cite{kim2021oncogenic}, and thus further zero in on the residues that we identified as domain $A_1$.
Residues 83-89, which we identified as part of domain $A_2$, correspond to a loop that precedes the helix $\alpha_3$ in KRAS and has been identified as an allosteric site involved in the modulation of nucleotide hydrolysis~\cite{jani2024insight,bery2019kras}.
Similarly, we identified residues 76-77 as domain $B_1$, which are located at one end of the switch II domain.
Interestingly, residues 24-28, which form domain $B_2$ and feature anti-correlated motion relative to domain $B_1$, are located at the beginning of the switch I domain, indicating a dynamic coupling between both conserved domains.

We note that our detection of residues in highly conserved regulatory domains of KRAS did not require prior any knowledge or long simulation trajectories. 
Our enhanced sampling simulations utilized anharmonic low-frequency vibrations as CVs that were detected in 20~ns simulations. 
While we averaged results over 20 independent enhanced sampling simulations, neither of them exceeded 100~ns of simulation time. 

\subsection{Projecting into Human-Readable Space}

\begin{figure}[ht!]
\begin{center}
\includegraphics[clip=true,width=0.6\linewidth]{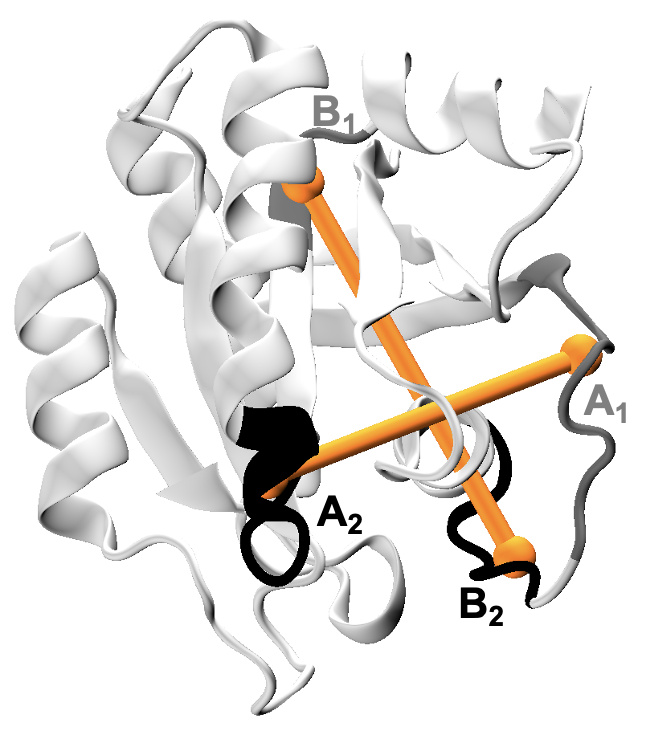}
\caption{
Visualization of the collectively moving domains $A_1$, $A_2$, $B_1$ and $B_2$ (gray and black) and their corresponding interdomain distances (orange) in the 3D structure of KRAS (cartoon representation).
}
\label{f:dom}
\end{center}
\end{figure}

Center-of-mass distances between collectively moving domains in the anti-correlated pairs $d_A=\left|A_1-A_2\right|$ and $d_B=\left|B_1-B_2\right|$ provide intuitive choices for CVs that are straightforward to interpret. 
Further, we expect them to capture the essential conformational changes of KRAS.
We visualized both pairs of domains and the vectors that connect them in Figure~\ref{f:dom}.

\begin{figure}[ht!]
\begin{center}
\includegraphics[width=1.0\linewidth]{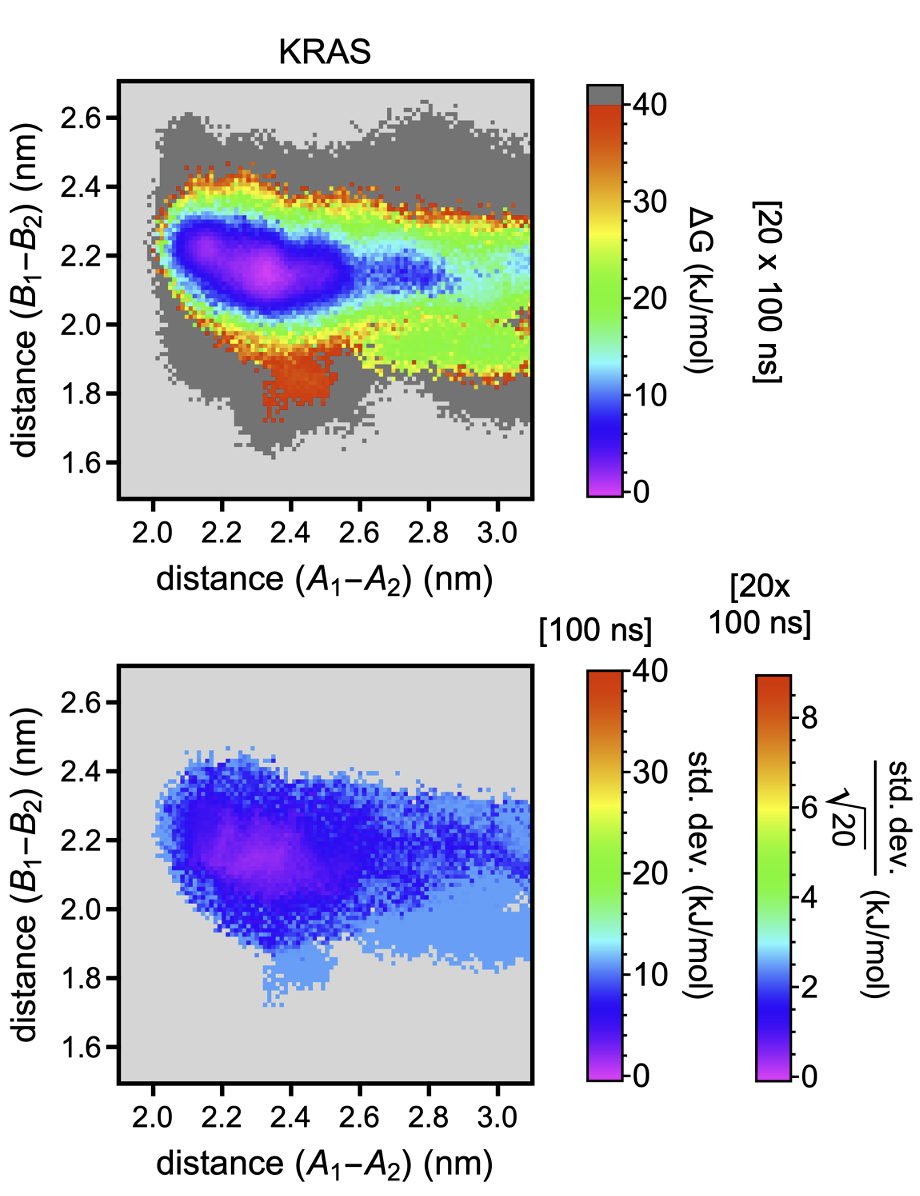}
\caption{
Conformational ensemble of KRAS as a function of the domain-domain distances $d_A=\left|A_1-A_2\right|$ and $d_B=\left|B_1-B_2\right|$.
The average free energy surface (top panel) is obtained after unbiasing 20 $\times$ 100~ns metadynamics trajectories, generating weighted histograms as a function of $d_A$ and $d_B$, averaging the corresponding probability distributions, and conversion into free energies.
We estimated the statistical accuracy of our free energy surface (bottom panel) based on the standard deviation of the histograms obtained from the 20 metadynamics trajectories. 
The uncertainties are reported both as standard deviations and errors of the mean.
}
\label{f:fes}
\end{center}
\end{figure}

We then used both domain-domain distances to construct free energy surfaces after unbiasing the metadynamics simulations. 
The results for KRAS are shown in Figure~\ref{f:fes} together with statistical errors obtained from standard deviation over the 20 independent simulations.
Analogous free energy surfaces for HEWL, HIV-1Pr, MCL-1 and RBP as a function of their respective domain-domain distances are shown in \1{Figure~S3 of the SI}.

A notable feature of the domain-domain distances that we extract from the DCCM is that correlations between them are minimal by construction. 
This can increase the information content compared to CVs that are inherently correlated and thus do not provide independent information on the conformational changes.
The latter is frequently observed for user-defined CVs used in the literature~\cite{costa2023,huang2018replica,ren2021unraveling,benabderrahmane2020insights}.
Telltale signs are, for example, 2D free energy surfaces with free energy minima exclusively along the diagonal.
At the same time, correlations between our domain pairs are only minimized but not forbidden, and we expect to see any mechanistic correlations in our analysis of the resulting free energy surfaces.

Movement along the distance $d_A$ can be expected to be highly collective due to the specifics of our selection algorithm.
This is not necessarily the case for motion along $d_B$. 
In both cases, the degree of collectivity or size of a collectively moving domain can be easily assessed through visualization of the rows/columns $a_1$, $a_2$, $b_1$ and $b_2$ as shown in Figure~\ref{f:corr}.
For example, the domains $A_1$ and $A_2$ in KRAS in Figure~\ref{f:corr} consist of roughly 10 amino acids, while the corresponding domains in HIV-1 Pr or RBP are substantially larger (see \1{Figure~S2 in the SI}).  

Collective motion does not strictly imply large amplitudes.
In principle, a collective vibration around a sharp minimum would also give rise to large correlations in the DCCM used here to identify the domains $A_1$ and $A_2$. 
In practice, however, if the underlying enhanced sampling simulations sample a collective motion of large amplitude, as is the case here, the $d_A$ distance can be expected to be associated with it.

For KRAS, we observe a free energy minimum elongated along $d_A$ that allows for an almost barrier-free motion of roughly 0.4~nm for distances between 2.1 and 2.5~nm. 
Distances of 2.7~nm or 3.0~nm are accessible with moderate free energy costs of 10~kJ/mol and 15~kJ/mol, respectively.
Movement along $d_B$ is more restricted overall and somewhat dependent on $d_A$.
For $d_A$ between 2.1 and 2.3~nm, changes in $d_B$ are essentially anti-correlated.
For $d_A$ between 2.3 and 2.6~nm, $d_B$ remains mostly fixed at 2.1~nm, while for larger $d_A$, $\left|B_1-B_2\right|$ can switch between two states with $d_B$ close to 1.9~nm and 2.1~nm separated by a small barrier. 
Thus, this 0.2~nm motion of the domains $B_1$ and $B_2$ depends on the preceding separation of the domains $A_1$ and $A_2$.

When analyzing the finer details of the free energy surface, it is necessary to consider the statistical uncertainties associated with our analysis. 
The lower panel of Figure~\ref{f:fes} shows the statistical errors of the free energy. 
The latter are determined by error propagation from standard deviations and errors of the mean of the probability distributions obtained after weighting the configurations in independent metadynamics simulation trajectories as defined in Eq.~\ref{e:prob}.  

The lowest free energy states are generally sampled most reliably, even in biased simulations, and thus feature the lowest statistical uncertainty. 
However, even states with free energies around +20~kJ/mol relative to the minimum feature statistical errors below 3~kJ/mol after averaging over 20 trajectories.
We note that, in unbiased simulations, the probability to sample such states is approximately 3000 times lower compared to the global free energy minimum.
Thus, consistent sampling of these states can be directly attributed to the enhanced sampling along the low-frequency vibrations employed as CVs.

\section{Conclusion}
\label{s:con}
Enhanced sampling simulations of proteins and other biomolecules are increasingly based on CVs that are tuned to accelerate conformational transitions but are less amenable to straightforward human interpretation.
Our own recent work combines longstanding concepts on the role of low-frequency vibrations in protein conformational dynamics\cite{brooks1983harmonic} with a new approach to isolate these vibrations from short dynamics simulations that does not rely on harmonic approximations\cite{sauer2023frequency,mondal2024exploring,sauer2025high,sauer2026fast}.
Other recent approaches utilize ML to construct CVs that are nonlinear functions of input features used to train the network\cite{ribeiro2018reweighted,wang2019past,mehdi2024enhanced,hanni2025data}.

However, interpreting simulation trajectories that are biased along complex CVs can be challenging. 
A simple visualization of a biased trajectory can be highly misleading because large amplitude motions may feature low thermodynamic weights.
Similarly, free energy surfaces as a function of complex CVs are non-trivial to read. 
With sufficient sampling, biased simulations can always be recast into an alternative set of CVs. 
However, even then it can be a challenge of its own to choose a set of CVs that simultaneously captures the important dynamics while being easy to interpret.

Characterization of conformational ensembles extracted from unbiased simulations initiated for different states on free energy surfaces~\cite{istomin2024festa} can provide clues on key structural changes. 
However, such procedures remain a manual and potentially ambiguous task.

In particular, for CVs derived from ML methods, the computational cost associated with the evaluation of neural network-based CVs can be a burden during biased simulations and analysis.
This can be circumvented with suitable surrogate variables that approximate the output of non-linear neural networks as simplified linear combinations of a subset of the input features.\cite{chatterjee2025acceleration}
Such surrogate variables can also simplify direct interpretation if a surrogate variable is dominated by a small number of input features. 
However, even a surrogate variable dominated by a single input feature can be hard to interpret in terms of a specific structural change if that input feature is a highly non-linear function of the system coordinates, {\em e.g.}, a solvent coordination number.

Here, we proposed a fully automated procedure that identifies the distances between collectively moving domains as human-readable CVs that capture the essential dynamics obtained from enhanced sampling simulations.
Our method uses a weighted analysis of dynamic cross correlations for biased simulation trajectories and a simple algorithm to extract the collectively moving domains.
We applied this approach to a recent set of enhanced sampling simulations with known conformational dynamics and readily identified key inter-domain motions for each system.
Notably, the entire simulation protocol, which uses anharmonic low-frequency vibrations sampled from short equilibrium simulations as CVs, does not require any prior knowledge of the system and thus can be fully automated.

We anticipate that such fully automated enhanced sampling protocols will play a critical role in the generation of extensive protein conformational ensembles, which not only expand our knowledge of specific proteins but also provide essential training data for ML predictors of protein dynamics \cite{bioemu}.

\section{Methods}
\label{s:methods}

The detailed protocols for the all-atom molecular dynamics simulations of all systems are described in Ref.~\citenum{sauer2026fast}, which also provides brief descriptions of FRESEAN mode analysis and well-tempered metadynamics simulations. 
In summary, we performed molecular dynamics simulations with GROMACS~2022.5
beginning with crystal structures from the following PDB entries: 5WCC (KRAS), 1HEL (HEWL), 1BVE (HIV-1 Protease), 3WIX (MCL-1), and 1DRJ (RBP). 
The selection of force fields for each system was based on prior studies in the literature to enable direct comparisons.
\cite{benabderrahmane2020insights,desimone2013characterization,huang2018replica,chen2021mutation,ren2021unraveling}

\begin{figure}[ht!]
\begin{center}
\includegraphics[clip=true,width=0.6\linewidth]{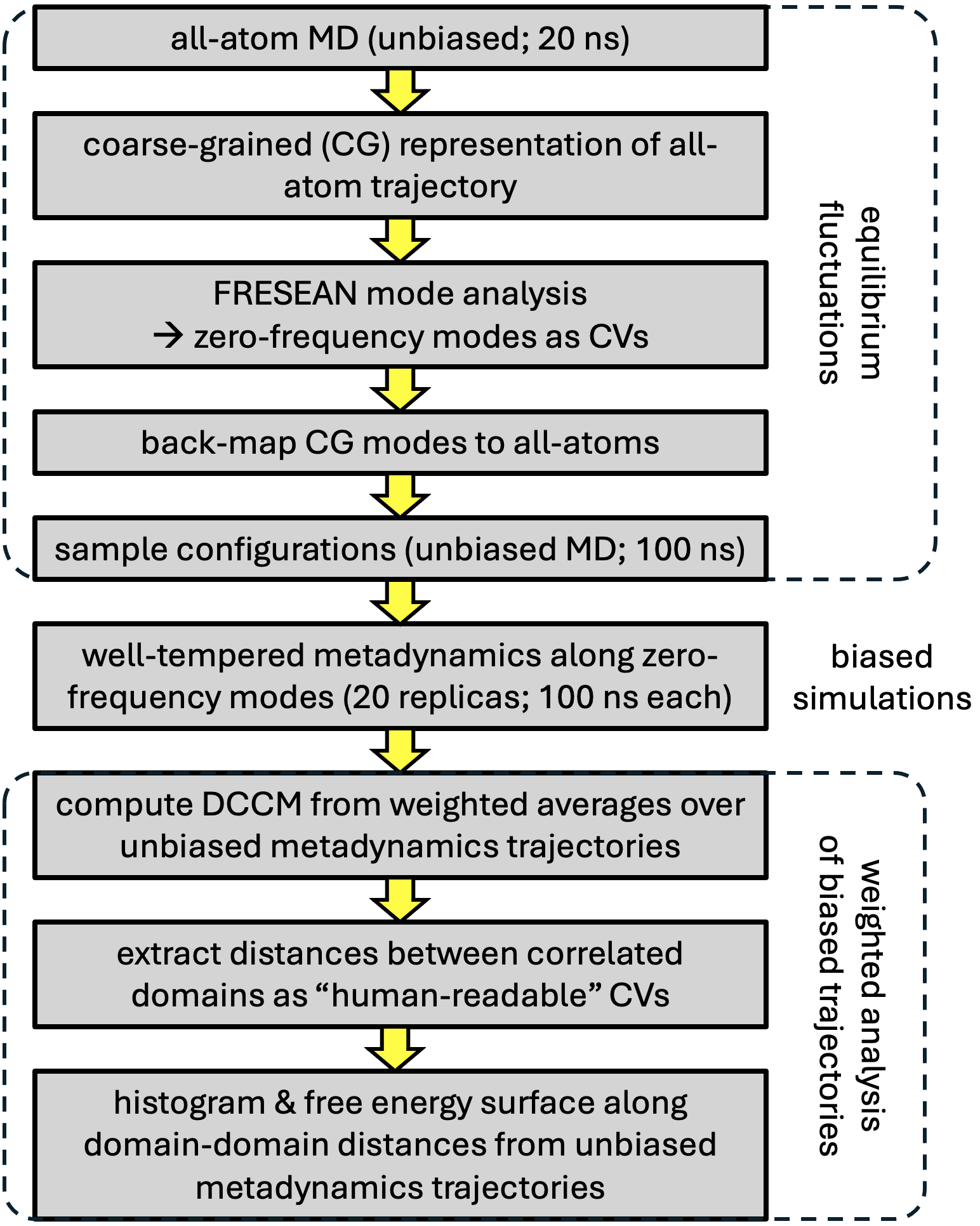}
\caption{
Enhanced sampling protocol with automatic generation of CVs suitable for enhanced sampling (zero-frequency modes detected from FRESEAN mode analysis) and projection on distances of anti-correlated domains as "human-readable" CVs.
}
\label{f:protocol}
\end{center}
\end{figure}

We adhered to the protocol outlined in Fig.~\ref{f:protocol}. 
After energy-minimzation and equilibration, we conducted unrestrained NPT simulations for 20~ns of each system during which coordinates and velocities were stored every 20~fs.
The all-atom trajectories of the protein were converted into a coarse-grained representation in which each amino acid (except for glycine) is represented by two beads (centers of mass and center of mass velocities of backbone and side chain atoms).~\cite{sauer2025high}
Glycine residues are presented by a single bead.
The coarse-grained trajectory is then aligned with a (coarse-grained) reference structure, where rotations are applied to both coordinates and velocities.
We then performed FRESEAN mode analysis by calculating a matrix of mass-weighted time velocity cross correlation functions for all coarse-grained degrees of freedom with a maximum correlation time of 2 ps.~\cite{sauer2023frequency} 
The matrix terms were time-symmetrized and transformed into the frequency domain with a Gaussian window function with a width of 10~\wn.

The two lowest-frequency vibrational modes (modes 7 and 8 at zero frequency) were chosen as collective variables (CVs) for enhanced sampling.\cite{sauer2026fast}

We performed an unbiased MD simulation for 100~ns and extracted 20 configurations (coordinates and velocities sampled every 5~ns) as starting configurations for separate replicas of well-tempered metadynamics simulations.~\cite{barducci2008well}
The metadynamics simulations were performed with the PLUMED 2.8.2 plugin in GROMACS~2022.5 using the zero-frequency modes as CVs after backmapping from the coarse-grained to an all-atom representation.
Each of metadynamics simulation was run for 100~ns and the resulting free energy surface (inverted cumulative biasing potential) was converted into an unbiased probability distribution prior to averaging and converting back to free energies.

\subsection*{Software Availability}
Source code and scripts for FRESEAN mode analysis, well-tempered metadynamics simulaitons with GROMACS and PLUMED, and the weighted DCCM analysis are available in our GitHub repository: https://github.com/HeydenLabASU-collab/FRESEAN-metadynamics and on Zenodo~\cite{FRESEANmetadynamics}.

\begin{acknowledgements}
This work is supported by the National Science Foundation (CHE-2154834) and the National Institute of General Medical Sciences (R01GM148622). The authors acknowledge Research Computing at Arizona State University for providing high performance computing resources that have contributed to the research results reported within this work.
\end{acknowledgements}


\begin{thebibliography}{57}%
\makeatletter
\providecommand \@ifxundefined [1]{%
 \@ifx{#1\undefined}
}%
\providecommand \@ifnum [1]{%
 \ifnum #1\expandafter \@firstoftwo
 \else \expandafter \@secondoftwo
 \fi
}%
\providecommand \@ifx [1]{%
 \ifx #1\expandafter \@firstoftwo
 \else \expandafter \@secondoftwo
 \fi
}%
\providecommand \natexlab [1]{#1}%
\providecommand \enquote  [1]{``#1''}%
\providecommand \bibnamefont  [1]{#1}%
\providecommand \bibfnamefont [1]{#1}%
\providecommand \citenamefont [1]{#1}%
\providecommand \href@noop [0]{\@secondoftwo}%
\providecommand \href [0]{\begingroup \@sanitize@url \@href}%
\providecommand \@href[1]{\@@startlink{#1}\@@href}%
\providecommand \@@href[1]{\endgroup#1\@@endlink}%
\providecommand \@sanitize@url [0]{\catcode `\\12\catcode `\$12\catcode
  `\&12\catcode `\#12\catcode `\^12\catcode `\_12\catcode `\%12\relax}%
\providecommand \@@startlink[1]{}%
\providecommand \@@endlink[0]{}%
\providecommand \url  [0]{\begingroup\@sanitize@url \@url }%
\providecommand \@url [1]{\endgroup\@href {#1}{\urlprefix }}%
\providecommand \urlprefix  [0]{URL }%
\providecommand \Eprint [0]{\href }%
\providecommand \doibase [0]{https://doi.org/}%
\providecommand \selectlanguage [0]{\@gobble}%
\providecommand \bibinfo  [0]{\@secondoftwo}%
\providecommand \bibfield  [0]{\@secondoftwo}%
\providecommand \translation [1]{[#1]}%
\providecommand \BibitemOpen [0]{}%
\providecommand \bibitemStop [0]{}%
\providecommand \bibitemNoStop [0]{.\EOS\space}%
\providecommand \EOS [0]{\spacefactor3000\relax}%
\providecommand \BibitemShut  [1]{\csname bibitem#1\endcsname}%
\let\auto@bib@innerbib\@empty
\bibitem [{\citenamefont {Henzler-Wildman}\ and\ \citenamefont
  {Kern}(2007)}]{henzler2007}%
  \BibitemOpen
  \bibfield  {author} {\bibinfo {author} {\bibfnamefont {K.}~\bibnamefont
  {Henzler-Wildman}}\ and\ \bibinfo {author} {\bibfnamefont {D.}~\bibnamefont
  {Kern}},\ }\bibfield  {title} {\bibinfo {title} {Dynamic personalities of
  proteins},\ }\href@noop {} {\bibfield  {journal} {\bibinfo  {journal}
  {Nature}\ }\textbf {\bibinfo {volume} {450}},\ \bibinfo {pages} {964}
  (\bibinfo {year} {2007})}\BibitemShut {NoStop}%
\bibitem [{\citenamefont {Wedemeyer}\ \emph {et~al.}(2002)\citenamefont
  {Wedemeyer}, \citenamefont {Welker},\ and\ \citenamefont
  {Scheraga}}]{wedemeyer2002}%
  \BibitemOpen
  \bibfield  {author} {\bibinfo {author} {\bibfnamefont {W.~J.}\ \bibnamefont
  {Wedemeyer}}, \bibinfo {author} {\bibfnamefont {E.}~\bibnamefont {Welker}},\
  and\ \bibinfo {author} {\bibfnamefont {H.~A.}\ \bibnamefont {Scheraga}},\
  }\bibfield  {title} {\bibinfo {title} {Proline cis- trans isomerization and
  protein folding},\ }\href@noop {} {\bibfield  {journal} {\bibinfo  {journal}
  {Biochemistry}\ }\textbf {\bibinfo {volume} {41}},\ \bibinfo {pages} {14637}
  (\bibinfo {year} {2002})}\BibitemShut {NoStop}%
\bibitem [{\citenamefont {Boehr}\ \emph {et~al.}(2006)\citenamefont {Boehr},
  \citenamefont {Dyson},\ and\ \citenamefont {Wright}}]{boehr2006}%
  \BibitemOpen
  \bibfield  {author} {\bibinfo {author} {\bibfnamefont {D.~D.}\ \bibnamefont
  {Boehr}}, \bibinfo {author} {\bibfnamefont {H.~J.}\ \bibnamefont {Dyson}},\
  and\ \bibinfo {author} {\bibfnamefont {P.~E.}\ \bibnamefont {Wright}},\
  }\bibfield  {title} {\bibinfo {title} {An nmr perspective on enzyme
  dynamics},\ }\href@noop {} {\bibfield  {journal} {\bibinfo  {journal} {Chem.
  Rev.}\ }\textbf {\bibinfo {volume} {106}},\ \bibinfo {pages} {3055} (\bibinfo
  {year} {2006})}\BibitemShut {NoStop}%
\bibitem [{\citenamefont {Karplus}\ and\ \citenamefont
  {McCammon}(2002)}]{karplus2002}%
  \BibitemOpen
  \bibfield  {author} {\bibinfo {author} {\bibfnamefont {M.}~\bibnamefont
  {Karplus}}\ and\ \bibinfo {author} {\bibfnamefont {J.~A.}\ \bibnamefont
  {McCammon}},\ }\bibfield  {title} {\bibinfo {title} {Molecular dynamics
  simulations of biomolecules},\ }\href@noop {} {\bibfield  {journal} {\bibinfo
   {journal} {Nat. Struct. Biol.}\ }\textbf {\bibinfo {volume} {9}},\ \bibinfo
  {pages} {646} (\bibinfo {year} {2002})}\BibitemShut {NoStop}%
\bibitem [{\citenamefont {Dror}\ \emph {et~al.}(2012)\citenamefont {Dror},
  \citenamefont {Dirks}, \citenamefont {Grossman}, \citenamefont {Xu},\ and\
  \citenamefont {Shaw}}]{dror2012}%
  \BibitemOpen
  \bibfield  {author} {\bibinfo {author} {\bibfnamefont {R.~O.}\ \bibnamefont
  {Dror}}, \bibinfo {author} {\bibfnamefont {R.~M.}\ \bibnamefont {Dirks}},
  \bibinfo {author} {\bibfnamefont {J.}~\bibnamefont {Grossman}}, \bibinfo
  {author} {\bibfnamefont {H.}~\bibnamefont {Xu}},\ and\ \bibinfo {author}
  {\bibfnamefont {D.~E.}\ \bibnamefont {Shaw}},\ }\bibfield  {title} {\bibinfo
  {title} {Biomolecular simulation: A computational microscope for molecular
  biology},\ }\href@noop {} {\bibfield  {journal} {\bibinfo  {journal} {Annu.
  Rev. Biophys.}\ }\textbf {\bibinfo {volume} {41}},\ \bibinfo {pages} {429}
  (\bibinfo {year} {2012})}\BibitemShut {NoStop}%
\bibitem [{\citenamefont {Shaw}\ \emph {et~al.}(2021)\citenamefont {Shaw},
  \citenamefont {Adams}, \citenamefont {Azaria}, \citenamefont {Bank},
  \citenamefont {Batson}, \citenamefont {Bell}, \citenamefont {Bergdorf},
  \citenamefont {Bhatt}, \citenamefont {Butts}, \citenamefont {Correia} \emph
  {et~al.}}]{shaw2021}%
  \BibitemOpen
  \bibfield  {author} {\bibinfo {author} {\bibfnamefont {D.~E.}\ \bibnamefont
  {Shaw}}, \bibinfo {author} {\bibfnamefont {P.~J.}\ \bibnamefont {Adams}},
  \bibinfo {author} {\bibfnamefont {A.}~\bibnamefont {Azaria}}, \bibinfo
  {author} {\bibfnamefont {J.~A.}\ \bibnamefont {Bank}}, \bibinfo {author}
  {\bibfnamefont {B.}~\bibnamefont {Batson}}, \bibinfo {author} {\bibfnamefont
  {A.}~\bibnamefont {Bell}}, \bibinfo {author} {\bibfnamefont {M.}~\bibnamefont
  {Bergdorf}}, \bibinfo {author} {\bibfnamefont {J.}~\bibnamefont {Bhatt}},
  \bibinfo {author} {\bibfnamefont {J.~A.}\ \bibnamefont {Butts}}, \bibinfo
  {author} {\bibfnamefont {T.}~\bibnamefont {Correia}}, \emph {et~al.},\
  }\bibfield  {title} {\bibinfo {title} {Anton 3: Twenty microseconds of
  molecular dynamics simulation before lunch},\ }in\ \href@noop {} {\emph
  {\bibinfo {booktitle} {Proceedings of the International Conference for High
  Performance Computing, Networking, Storage and Analysis}}}\ (\bibinfo {year}
  {2021})\ pp.\ \bibinfo {pages} {1--11}\BibitemShut {NoStop}%
\bibitem [{\citenamefont {Ayaz}\ \emph {et~al.}(2023)\citenamefont {Ayaz},
  \citenamefont {Lyczek}, \citenamefont {Paung}, \citenamefont {Mingione},
  \citenamefont {Iacob}, \citenamefont {de~Waal}, \citenamefont {Engen},
  \citenamefont {Seeliger}, \citenamefont {Shan},\ and\ \citenamefont
  {Shaw}}]{ayaz2023}%
  \BibitemOpen
  \bibfield  {author} {\bibinfo {author} {\bibfnamefont {P.}~\bibnamefont
  {Ayaz}}, \bibinfo {author} {\bibfnamefont {A.}~\bibnamefont {Lyczek}},
  \bibinfo {author} {\bibfnamefont {Y.}~\bibnamefont {Paung}}, \bibinfo
  {author} {\bibfnamefont {V.~R.}\ \bibnamefont {Mingione}}, \bibinfo {author}
  {\bibfnamefont {R.~E.}\ \bibnamefont {Iacob}}, \bibinfo {author}
  {\bibfnamefont {P.~W.}\ \bibnamefont {de~Waal}}, \bibinfo {author}
  {\bibfnamefont {J.~R.}\ \bibnamefont {Engen}}, \bibinfo {author}
  {\bibfnamefont {M.~A.}\ \bibnamefont {Seeliger}}, \bibinfo {author}
  {\bibfnamefont {Y.}~\bibnamefont {Shan}},\ and\ \bibinfo {author}
  {\bibfnamefont {D.~E.}\ \bibnamefont {Shaw}},\ }\bibfield  {title} {\bibinfo
  {title} {Structural mechanism of a drug-binding process involving a large
  conformational change of the protein target},\ }\href@noop {} {\bibfield
  {journal} {\bibinfo  {journal} {Nat. Commun.}\ }\textbf {\bibinfo {volume}
  {14}},\ \bibinfo {pages} {1885} (\bibinfo {year} {2023})}\BibitemShut
  {NoStop}%
\bibitem [{\citenamefont {Greisman}\ \emph {et~al.}(2023)\citenamefont
  {Greisman}, \citenamefont {Willmore}, \citenamefont {Yeh}, \citenamefont
  {Giordanetto}, \citenamefont {Shahamadtar}, \citenamefont {Nisonoff},
  \citenamefont {Maragakis},\ and\ \citenamefont {Shaw}}]{greisman2023}%
  \BibitemOpen
  \bibfield  {author} {\bibinfo {author} {\bibfnamefont {J.~B.}\ \bibnamefont
  {Greisman}}, \bibinfo {author} {\bibfnamefont {L.}~\bibnamefont {Willmore}},
  \bibinfo {author} {\bibfnamefont {C.~Y.}\ \bibnamefont {Yeh}}, \bibinfo
  {author} {\bibfnamefont {F.}~\bibnamefont {Giordanetto}}, \bibinfo {author}
  {\bibfnamefont {S.}~\bibnamefont {Shahamadtar}}, \bibinfo {author}
  {\bibfnamefont {H.}~\bibnamefont {Nisonoff}}, \bibinfo {author}
  {\bibfnamefont {P.}~\bibnamefont {Maragakis}},\ and\ \bibinfo {author}
  {\bibfnamefont {D.~E.}\ \bibnamefont {Shaw}},\ }\bibfield  {title} {\bibinfo
  {title} {Discovery and validation of the binding poses of allosteric fragment
  hits to protein tyrosine phosphatase 1b: From molecular dynamics simulations
  to x-ray crystallography},\ }\href@noop {} {\bibfield  {journal} {\bibinfo
  {journal} {J. Chem. Inf. and Model.}\ }\textbf {\bibinfo {volume} {63}},\
  \bibinfo {pages} {2644} (\bibinfo {year} {2023})}\BibitemShut {NoStop}%
\bibitem [{\citenamefont {Lewis}\ \emph {et~al.}(2025)\citenamefont {Lewis},
  \citenamefont {Hempel}, \citenamefont {Jim{\'e}nez-Luna}, \citenamefont
  {Gastegger}, \citenamefont {Xie}, \citenamefont {Foong}, \citenamefont
  {Satorras}, \citenamefont {Abdin}, \citenamefont {Veeling}, \citenamefont
  {Zaporozhets}, \citenamefont {Chen}, \citenamefont {Yang}, \citenamefont
  {Foster}, \citenamefont {Schneuing}, \citenamefont {Nigam}, \citenamefont
  {Barbero}, \citenamefont {Stimper}, \citenamefont {Campbell}, \citenamefont
  {Yim}, \citenamefont {Lienen}, \citenamefont {Shi}, \citenamefont {Zheng},
  \citenamefont {Schulz}, \citenamefont {Munir}, \citenamefont {Sordillo},
  \citenamefont {Tomioka}, \citenamefont {Clementi},\ and\ \citenamefont
  {No{\'e}}}]{bioemu}%
  \BibitemOpen
  \bibfield  {author} {\bibinfo {author} {\bibfnamefont {S.}~\bibnamefont
  {Lewis}}, \bibinfo {author} {\bibfnamefont {T.}~\bibnamefont {Hempel}},
  \bibinfo {author} {\bibfnamefont {J.}~\bibnamefont {Jim{\'e}nez-Luna}},
  \bibinfo {author} {\bibfnamefont {M.}~\bibnamefont {Gastegger}}, \bibinfo
  {author} {\bibfnamefont {Y.}~\bibnamefont {Xie}}, \bibinfo {author}
  {\bibfnamefont {A.~Y.}\ \bibnamefont {Foong}}, \bibinfo {author}
  {\bibfnamefont {V.~G.}\ \bibnamefont {Satorras}}, \bibinfo {author}
  {\bibfnamefont {O.}~\bibnamefont {Abdin}}, \bibinfo {author} {\bibfnamefont
  {B.~S.}\ \bibnamefont {Veeling}}, \bibinfo {author} {\bibfnamefont
  {I.}~\bibnamefont {Zaporozhets}}, \bibinfo {author} {\bibfnamefont
  {Y.}~\bibnamefont {Chen}}, \bibinfo {author} {\bibfnamefont {S.}~\bibnamefont
  {Yang}}, \bibinfo {author} {\bibfnamefont {A.~E.}\ \bibnamefont {Foster}},
  \bibinfo {author} {\bibfnamefont {A.}~\bibnamefont {Schneuing}}, \bibinfo
  {author} {\bibfnamefont {J.}~\bibnamefont {Nigam}}, \bibinfo {author}
  {\bibfnamefont {F.}~\bibnamefont {Barbero}}, \bibinfo {author} {\bibfnamefont
  {V.}~\bibnamefont {Stimper}}, \bibinfo {author} {\bibfnamefont
  {A.}~\bibnamefont {Campbell}}, \bibinfo {author} {\bibfnamefont
  {J.}~\bibnamefont {Yim}}, \bibinfo {author} {\bibfnamefont {M.}~\bibnamefont
  {Lienen}}, \bibinfo {author} {\bibfnamefont {Y.}~\bibnamefont {Shi}},
  \bibinfo {author} {\bibfnamefont {S.}~\bibnamefont {Zheng}}, \bibinfo
  {author} {\bibfnamefont {H.}~\bibnamefont {Schulz}}, \bibinfo {author}
  {\bibfnamefont {U.}~\bibnamefont {Munir}}, \bibinfo {author} {\bibfnamefont
  {R.}~\bibnamefont {Sordillo}}, \bibinfo {author} {\bibfnamefont
  {R.}~\bibnamefont {Tomioka}}, \bibinfo {author} {\bibfnamefont
  {C.}~\bibnamefont {Clementi}},\ and\ \bibinfo {author} {\bibfnamefont
  {F.}~\bibnamefont {No{\'e}}},\ }\bibfield  {title} {\bibinfo {title}
  {Scalable emulation of protein equilibrium ensembles with generative deep
  learning},\ }\href {https://doi.org/10.1126/science.adv9817} {\bibfield
  {journal} {\bibinfo  {journal} {Science}\ }\textbf {\bibinfo {volume}
  {389}},\ \bibinfo {pages} {eadv9817} (\bibinfo {year} {2025})}\BibitemShut
  {NoStop}%
\bibitem [{\citenamefont {Bernardi}\ \emph {et~al.}(2015)\citenamefont
  {Bernardi}, \citenamefont {Melo},\ and\ \citenamefont
  {Schulten}}]{bernardi2015}%
  \BibitemOpen
  \bibfield  {author} {\bibinfo {author} {\bibfnamefont {R.~C.}\ \bibnamefont
  {Bernardi}}, \bibinfo {author} {\bibfnamefont {M.~C.}\ \bibnamefont {Melo}},\
  and\ \bibinfo {author} {\bibfnamefont {K.}~\bibnamefont {Schulten}},\
  }\bibfield  {title} {\bibinfo {title} {Enhanced sampling techniques in
  molecular dynamics simulations of biological systems},\ }\href@noop {}
  {\bibfield  {journal} {\bibinfo  {journal} {Biochim. Biophys. Acta, Gen.
  Subj.}\ }\textbf {\bibinfo {volume} {1850}},\ \bibinfo {pages} {872}
  (\bibinfo {year} {2015})}\BibitemShut {NoStop}%
\bibitem [{\citenamefont {Mitsutake}\ \emph {et~al.}(2001)\citenamefont
  {Mitsutake}, \citenamefont {Sugita},\ and\ \citenamefont
  {Okamoto}}]{mitsutake2001}%
  \BibitemOpen
  \bibfield  {author} {\bibinfo {author} {\bibfnamefont {A.}~\bibnamefont
  {Mitsutake}}, \bibinfo {author} {\bibfnamefont {Y.}~\bibnamefont {Sugita}},\
  and\ \bibinfo {author} {\bibfnamefont {Y.}~\bibnamefont {Okamoto}},\
  }\bibfield  {title} {\bibinfo {title} {Generalized-ensemble algorithms for
  molecular simulations of biopolymers},\ }\href@noop {} {\bibfield  {journal}
  {\bibinfo  {journal} {Biomolecules}\ }\textbf {\bibinfo {volume} {60}},\
  \bibinfo {pages} {96} (\bibinfo {year} {2001})}\BibitemShut {NoStop}%
\bibitem [{\citenamefont {Torrie}\ and\ \citenamefont
  {Valleau}(1977)}]{torrie1977}%
  \BibitemOpen
  \bibfield  {author} {\bibinfo {author} {\bibfnamefont {G.~M.}\ \bibnamefont
  {Torrie}}\ and\ \bibinfo {author} {\bibfnamefont {J.~P.}\ \bibnamefont
  {Valleau}},\ }\bibfield  {title} {\bibinfo {title} {Nonphysical sampling
  distributions in monte carlo free-energy estimation: Umbrella sampling},\
  }\href@noop {} {\bibfield  {journal} {\bibinfo  {journal} {J. Comput. Phys.}\
  }\textbf {\bibinfo {volume} {23}},\ \bibinfo {pages} {187} (\bibinfo {year}
  {1977})}\BibitemShut {NoStop}%
\bibitem [{\citenamefont {Isralewitz}\ \emph {et~al.}(2001)\citenamefont
  {Isralewitz}, \citenamefont {Baudry}, \citenamefont {Gullingsrud},
  \citenamefont {Kosztin},\ and\ \citenamefont {Schulten}}]{isralewitz2001}%
  \BibitemOpen
  \bibfield  {author} {\bibinfo {author} {\bibfnamefont {B.}~\bibnamefont
  {Isralewitz}}, \bibinfo {author} {\bibfnamefont {J.}~\bibnamefont {Baudry}},
  \bibinfo {author} {\bibfnamefont {J.}~\bibnamefont {Gullingsrud}}, \bibinfo
  {author} {\bibfnamefont {D.}~\bibnamefont {Kosztin}},\ and\ \bibinfo {author}
  {\bibfnamefont {K.}~\bibnamefont {Schulten}},\ }\bibfield  {title} {\bibinfo
  {title} {Steered molecular dynamics investigations of protein function},\
  }\href@noop {} {\bibfield  {journal} {\bibinfo  {journal} {J. Mol. Graph.
  Model.}\ }\textbf {\bibinfo {volume} {19}},\ \bibinfo {pages} {13} (\bibinfo
  {year} {2001})}\BibitemShut {NoStop}%
\bibitem [{\citenamefont {Schlitter}\ \emph {et~al.}(1994)\citenamefont
  {Schlitter}, \citenamefont {Engels},\ and\ \citenamefont
  {Kr{\"u}ger}}]{schlitter1994}%
  \BibitemOpen
  \bibfield  {author} {\bibinfo {author} {\bibfnamefont {J.}~\bibnamefont
  {Schlitter}}, \bibinfo {author} {\bibfnamefont {M.}~\bibnamefont {Engels}},\
  and\ \bibinfo {author} {\bibfnamefont {P.}~\bibnamefont {Kr{\"u}ger}},\
  }\bibfield  {title} {\bibinfo {title} {Targeted molecular dynamics: a new
  approach for searching pathways of conformational transitions},\ }\href@noop
  {} {\bibfield  {journal} {\bibinfo  {journal} {J. Mol. Graph.}\ }\textbf
  {\bibinfo {volume} {12}},\ \bibinfo {pages} {84} (\bibinfo {year}
  {1994})}\BibitemShut {NoStop}%
\bibitem [{\citenamefont {H{\'e}nin}\ and\ \citenamefont
  {Chipot}(2004)}]{henin2004}%
  \BibitemOpen
  \bibfield  {author} {\bibinfo {author} {\bibfnamefont {J.}~\bibnamefont
  {H{\'e}nin}}\ and\ \bibinfo {author} {\bibfnamefont {C.}~\bibnamefont
  {Chipot}},\ }\bibfield  {title} {\bibinfo {title} {Overcoming free energy
  barriers using unconstrained molecular dynamics simulations},\ }\href@noop {}
  {\bibfield  {journal} {\bibinfo  {journal} {J. Chem. Phys.}\ }\textbf
  {\bibinfo {volume} {121}},\ \bibinfo {pages} {2904} (\bibinfo {year}
  {2004})}\BibitemShut {NoStop}%
\bibitem [{\citenamefont {Laio}\ and\ \citenamefont
  {Parrinello}(2002)}]{laio2002}%
  \BibitemOpen
  \bibfield  {author} {\bibinfo {author} {\bibfnamefont {A.}~\bibnamefont
  {Laio}}\ and\ \bibinfo {author} {\bibfnamefont {M.}~\bibnamefont
  {Parrinello}},\ }\bibfield  {title} {\bibinfo {title} {Escaping free-energy
  minima},\ }\href@noop {} {\bibfield  {journal} {\bibinfo  {journal} {Proc.
  Natl. Acad. Sci. U.S.A}\ }\textbf {\bibinfo {volume} {99}},\ \bibinfo {pages}
  {12562} (\bibinfo {year} {2002})}\BibitemShut {NoStop}%
\bibitem [{\citenamefont {Sugita}\ and\ \citenamefont
  {Okamoto}(1999)}]{sugita1999}%
  \BibitemOpen
  \bibfield  {author} {\bibinfo {author} {\bibfnamefont {Y.}~\bibnamefont
  {Sugita}}\ and\ \bibinfo {author} {\bibfnamefont {Y.}~\bibnamefont
  {Okamoto}},\ }\bibfield  {title} {\bibinfo {title} {Replica-exchange
  molecular dynamics method for protein folding},\ }\href@noop {} {\bibfield
  {journal} {\bibinfo  {journal} {Chem. Phys. Lett.}\ }\textbf {\bibinfo
  {volume} {314}},\ \bibinfo {pages} {141} (\bibinfo {year}
  {1999})}\BibitemShut {NoStop}%
\bibitem [{\citenamefont {Hamelberg}\ \emph {et~al.}(2004)\citenamefont
  {Hamelberg}, \citenamefont {Mongan},\ and\ \citenamefont
  {McCammon}}]{hamelberg2004}%
  \BibitemOpen
  \bibfield  {author} {\bibinfo {author} {\bibfnamefont {D.}~\bibnamefont
  {Hamelberg}}, \bibinfo {author} {\bibfnamefont {J.}~\bibnamefont {Mongan}},\
  and\ \bibinfo {author} {\bibfnamefont {J.~A.}\ \bibnamefont {McCammon}},\
  }\bibfield  {title} {\bibinfo {title} {Accelerated molecular dynamics: a
  promising and efficient simulation method for biomolecules},\ }\href@noop {}
  {\bibfield  {journal} {\bibinfo  {journal} {J. Chem. Phys.}\ }\textbf
  {\bibinfo {volume} {120}},\ \bibinfo {pages} {11919} (\bibinfo {year}
  {2004})}\BibitemShut {NoStop}%
\bibitem [{\citenamefont {Mehdi}\ \emph {et~al.}(2024)\citenamefont {Mehdi},
  \citenamefont {Smith}, \citenamefont {Herron}, \citenamefont {Zou},\ and\
  \citenamefont {Tiwary}}]{mehdi2024enhanced}%
  \BibitemOpen
  \bibfield  {author} {\bibinfo {author} {\bibfnamefont {S.}~\bibnamefont
  {Mehdi}}, \bibinfo {author} {\bibfnamefont {Z.}~\bibnamefont {Smith}},
  \bibinfo {author} {\bibfnamefont {L.}~\bibnamefont {Herron}}, \bibinfo
  {author} {\bibfnamefont {Z.}~\bibnamefont {Zou}},\ and\ \bibinfo {author}
  {\bibfnamefont {P.}~\bibnamefont {Tiwary}},\ }\bibfield  {title} {\bibinfo
  {title} {Enhanced sampling with machine learning},\ }\href@noop {} {\bibfield
   {journal} {\bibinfo  {journal} {Ann. Rev. Phys. Chem.}\ }\textbf {\bibinfo
  {volume} {75}} (\bibinfo {year} {2024})}\BibitemShut {NoStop}%
\bibitem [{\citenamefont {Go}\ \emph {et~al.}(1983)\citenamefont {Go},
  \citenamefont {Noguti},\ and\ \citenamefont {Nishikawa}}]{go1983}%
  \BibitemOpen
  \bibfield  {author} {\bibinfo {author} {\bibfnamefont {N.}~\bibnamefont
  {Go}}, \bibinfo {author} {\bibfnamefont {T.}~\bibnamefont {Noguti}},\ and\
  \bibinfo {author} {\bibfnamefont {T.}~\bibnamefont {Nishikawa}},\ }\bibfield
  {title} {\bibinfo {title} {Dynamics of a small globular protein in terms of
  low-frequency vibrational modes.},\ }\href@noop {} {\bibfield  {journal}
  {\bibinfo  {journal} {Proc. Natl. Acad. Sci. U.S.A}\ }\textbf {\bibinfo
  {volume} {80}},\ \bibinfo {pages} {3696} (\bibinfo {year}
  {1983})}\BibitemShut {NoStop}%
\bibitem [{\citenamefont {Aalten}\ \emph {et~al.}(1995)\citenamefont {Aalten},
  \citenamefont {Findlay}, \citenamefont {Amadei},\ and\ \citenamefont
  {Berendsen}}]{aalten1995}%
  \BibitemOpen
  \bibfield  {author} {\bibinfo {author} {\bibfnamefont {D.~v.}\ \bibnamefont
  {Aalten}}, \bibinfo {author} {\bibfnamefont {J.}~\bibnamefont {Findlay}},
  \bibinfo {author} {\bibfnamefont {A.}~\bibnamefont {Amadei}},\ and\ \bibinfo
  {author} {\bibfnamefont {H.}~\bibnamefont {Berendsen}},\ }\bibfield  {title}
  {\bibinfo {title} {Essential dynamics of the cellular retinol-binding protein
  evidence for ligand-induced conformational changes},\ }\href@noop {}
  {\bibfield  {journal} {\bibinfo  {journal} {Protein Eng. Des. Sel.}\ }\textbf
  {\bibinfo {volume} {8}},\ \bibinfo {pages} {1129} (\bibinfo {year}
  {1995})}\BibitemShut {NoStop}%
\bibitem [{\citenamefont {Hayward}\ \emph {et~al.}(1995)\citenamefont
  {Hayward}, \citenamefont {Kitao},\ and\ \citenamefont
  {G{\=o}}}]{hayward1995}%
  \BibitemOpen
  \bibfield  {author} {\bibinfo {author} {\bibfnamefont {S.}~\bibnamefont
  {Hayward}}, \bibinfo {author} {\bibfnamefont {A.}~\bibnamefont {Kitao}},\
  and\ \bibinfo {author} {\bibfnamefont {N.}~\bibnamefont {G{\=o}}},\
  }\bibfield  {title} {\bibinfo {title} {Harmonicity and anharmonicity in
  protein dynamics: a normal mode analysis and principal component analysis},\
  }\href@noop {} {\bibfield  {journal} {\bibinfo  {journal} {Proteins. Struct.
  Funct. Bioinf.}\ }\textbf {\bibinfo {volume} {23}},\ \bibinfo {pages} {177}
  (\bibinfo {year} {1995})}\BibitemShut {NoStop}%
\bibitem [{\citenamefont {Ma}(2005)}]{ma2005}%
  \BibitemOpen
  \bibfield  {author} {\bibinfo {author} {\bibfnamefont {J.}~\bibnamefont
  {Ma}},\ }\bibfield  {title} {\bibinfo {title} {Usefulness and limitations of
  normal mode analysis in modeling dynamics of biomolecular complexes},\
  }\href@noop {} {\bibfield  {journal} {\bibinfo  {journal} {Structure}\
  }\textbf {\bibinfo {volume} {13}},\ \bibinfo {pages} {373} (\bibinfo {year}
  {2005})}\BibitemShut {NoStop}%
\bibitem [{\citenamefont {Sauer}\ and\ \citenamefont
  {Heyden}(2023)}]{sauer2023frequency}%
  \BibitemOpen
  \bibfield  {author} {\bibinfo {author} {\bibfnamefont {M.~A.}\ \bibnamefont
  {Sauer}}\ and\ \bibinfo {author} {\bibfnamefont {M.}~\bibnamefont {Heyden}},\
  }\bibfield  {title} {\bibinfo {title} {Frequency-selective anharmonic mode
  analysis of thermally excited vibrations in proteins},\ }\href@noop {}
  {\bibfield  {journal} {\bibinfo  {journal} {J. Chem. Theory Comput.}\
  }\textbf {\bibinfo {volume} {19}},\ \bibinfo {pages} {5481} (\bibinfo {year}
  {2023})}\BibitemShut {NoStop}%
\bibitem [{\citenamefont {Garc{\'\i}a}(1992)}]{garcia1992large}%
  \BibitemOpen
  \bibfield  {author} {\bibinfo {author} {\bibfnamefont {A.~E.}\ \bibnamefont
  {Garc{\'\i}a}},\ }\bibfield  {title} {\bibinfo {title} {Large-amplitude
  nonlinear motions in proteins},\ }\href@noop {} {\bibfield  {journal}
  {\bibinfo  {journal} {Phys. Rev. Lett.}\ }\textbf {\bibinfo {volume} {68}},\
  \bibinfo {pages} {2696} (\bibinfo {year} {1992})}\BibitemShut {NoStop}%
\bibitem [{\citenamefont {Amadei}\ \emph {et~al.}(1993)\citenamefont {Amadei},
  \citenamefont {Linssen},\ and\ \citenamefont
  {Berendsen}}]{amadei1993essential}%
  \BibitemOpen
  \bibfield  {author} {\bibinfo {author} {\bibfnamefont {A.}~\bibnamefont
  {Amadei}}, \bibinfo {author} {\bibfnamefont {A.~B.}\ \bibnamefont
  {Linssen}},\ and\ \bibinfo {author} {\bibfnamefont {H.~J.}\ \bibnamefont
  {Berendsen}},\ }\bibfield  {title} {\bibinfo {title} {Essential dynamics of
  proteins},\ }\href@noop {} {\bibfield  {journal} {\bibinfo  {journal}
  {Proteins Struct. Funct. Bioinf.}\ }\textbf {\bibinfo {volume} {17}},\
  \bibinfo {pages} {412} (\bibinfo {year} {1993})}\BibitemShut {NoStop}%
\bibitem [{\citenamefont {Naritomi}\ and\ \citenamefont
  {Fuchigami}(2013)}]{naritomi2013}%
  \BibitemOpen
  \bibfield  {author} {\bibinfo {author} {\bibfnamefont {Y.}~\bibnamefont
  {Naritomi}}\ and\ \bibinfo {author} {\bibfnamefont {S.}~\bibnamefont
  {Fuchigami}},\ }\bibfield  {title} {\bibinfo {title} {Slow dynamics of a
  protein backbone in molecular dynamics simulation revealed by time-structure
  based independent component analysis},\ }\href@noop {} {\bibfield  {journal}
  {\bibinfo  {journal} {J. Chem. Phys.}\ }\textbf {\bibinfo {volume} {139}}
  (\bibinfo {year} {2013})}\BibitemShut {NoStop}%
\bibitem [{\citenamefont {No{\'e}}\ and\ \citenamefont
  {Clementi}(2015)}]{noe2015kinetic}%
  \BibitemOpen
  \bibfield  {author} {\bibinfo {author} {\bibfnamefont {F.}~\bibnamefont
  {No{\'e}}}\ and\ \bibinfo {author} {\bibfnamefont {C.}~\bibnamefont
  {Clementi}},\ }\bibfield  {title} {\bibinfo {title} {Kinetic distance and
  kinetic maps from molecular dynamics simulation},\ }\href@noop {} {\bibfield
  {journal} {\bibinfo  {journal} {J. Chem. Theory Comput.}\ }\textbf {\bibinfo
  {volume} {11}},\ \bibinfo {pages} {5002} (\bibinfo {year}
  {2015})}\BibitemShut {NoStop}%
\bibitem [{\citenamefont {Sidky}\ \emph {et~al.}(2020)\citenamefont {Sidky},
  \citenamefont {Chen},\ and\ \citenamefont {Ferguson}}]{sidky2020machine}%
  \BibitemOpen
  \bibfield  {author} {\bibinfo {author} {\bibfnamefont {H.}~\bibnamefont
  {Sidky}}, \bibinfo {author} {\bibfnamefont {W.}~\bibnamefont {Chen}},\ and\
  \bibinfo {author} {\bibfnamefont {A.~L.}\ \bibnamefont {Ferguson}},\
  }\bibfield  {title} {\bibinfo {title} {Machine learning for collective
  variable discovery and enhanced sampling in biomolecular simulation},\
  }\href@noop {} {\bibfield  {journal} {\bibinfo  {journal} {Mol. Phys.}\
  }\textbf {\bibinfo {volume} {118}},\ \bibinfo {pages} {e1737742} (\bibinfo
  {year} {2020})}\BibitemShut {NoStop}%
\bibitem [{\citenamefont {Wang}\ \emph {et~al.}(2020)\citenamefont {Wang},
  \citenamefont {Ribeiro},\ and\ \citenamefont {Tiwary}}]{wang2020machine}%
  \BibitemOpen
  \bibfield  {author} {\bibinfo {author} {\bibfnamefont {Y.}~\bibnamefont
  {Wang}}, \bibinfo {author} {\bibfnamefont {J.~M.~L.}\ \bibnamefont
  {Ribeiro}},\ and\ \bibinfo {author} {\bibfnamefont {P.}~\bibnamefont
  {Tiwary}},\ }\bibfield  {title} {\bibinfo {title} {Machine learning
  approaches for analyzing and enhancing molecular dynamics simulations},\
  }\href@noop {} {\bibfield  {journal} {\bibinfo  {journal} {Curr. Opin.
  Struct. Biol.}\ }\textbf {\bibinfo {volume} {61}},\ \bibinfo {pages} {139}
  (\bibinfo {year} {2020})}\BibitemShut {NoStop}%
\bibitem [{\citenamefont {No{\'e}}\ \emph {et~al.}(2020)\citenamefont
  {No{\'e}}, \citenamefont {Tkatchenko}, \citenamefont {M{\"u}ller},\ and\
  \citenamefont {Clementi}}]{noe2020machine}%
  \BibitemOpen
  \bibfield  {author} {\bibinfo {author} {\bibfnamefont {F.}~\bibnamefont
  {No{\'e}}}, \bibinfo {author} {\bibfnamefont {A.}~\bibnamefont {Tkatchenko}},
  \bibinfo {author} {\bibfnamefont {K.-R.}\ \bibnamefont {M{\"u}ller}},\ and\
  \bibinfo {author} {\bibfnamefont {C.}~\bibnamefont {Clementi}},\ }\bibfield
  {title} {\bibinfo {title} {Machine learning for molecular simulation},\
  }\href@noop {} {\bibfield  {journal} {\bibinfo  {journal} {Annu. Rev. Phys.
  Chem.}\ }\textbf {\bibinfo {volume} {71}},\ \bibinfo {pages} {361} (\bibinfo
  {year} {2020})}\BibitemShut {NoStop}%
\bibitem [{\citenamefont {Hanni}\ and\ \citenamefont
  {Ray}(2025)}]{hanni2025data}%
  \BibitemOpen
  \bibfield  {author} {\bibinfo {author} {\bibfnamefont {J.}~\bibnamefont
  {Hanni}}\ and\ \bibinfo {author} {\bibfnamefont {D.}~\bibnamefont {Ray}},\
  }\bibfield  {title} {\bibinfo {title} {Data efficient learning of molecular
  slow modes from nonequilibrium metadynamics},\ }\href@noop {} {\bibfield
  {journal} {\bibinfo  {journal} {J. Chem. Phys.}\ }\textbf {\bibinfo {volume}
  {162}} (\bibinfo {year} {2025})}\BibitemShut {NoStop}%
\bibitem [{\citenamefont {Fu}\ \emph {et~al.}(2024)\citenamefont {Fu},
  \citenamefont {Bian}, \citenamefont {Shao},\ and\ \citenamefont
  {Cai}}]{fu2024collective}%
  \BibitemOpen
  \bibfield  {author} {\bibinfo {author} {\bibfnamefont {H.}~\bibnamefont
  {Fu}}, \bibinfo {author} {\bibfnamefont {H.}~\bibnamefont {Bian}}, \bibinfo
  {author} {\bibfnamefont {X.}~\bibnamefont {Shao}},\ and\ \bibinfo {author}
  {\bibfnamefont {W.}~\bibnamefont {Cai}},\ }\bibfield  {title} {\bibinfo
  {title} {Collective variable-based enhanced sampling: From human learning to
  machine learning},\ }\href@noop {} {\bibfield  {journal} {\bibinfo  {journal}
  {J. Phys. Chem. Lett.}\ }\textbf {\bibinfo {volume} {15}},\ \bibinfo {pages}
  {1774} (\bibinfo {year} {2024})}\BibitemShut {NoStop}%
\bibitem [{\citenamefont {Fr{\"o}hlking}\ \emph {et~al.}(2024)\citenamefont
  {Fr{\"o}hlking}, \citenamefont {Bonati}, \citenamefont {Rizzi},\ and\
  \citenamefont {Gervasio}}]{frohlking2024deep}%
  \BibitemOpen
  \bibfield  {author} {\bibinfo {author} {\bibfnamefont {T.}~\bibnamefont
  {Fr{\"o}hlking}}, \bibinfo {author} {\bibfnamefont {L.}~\bibnamefont
  {Bonati}}, \bibinfo {author} {\bibfnamefont {V.}~\bibnamefont {Rizzi}},\ and\
  \bibinfo {author} {\bibfnamefont {F.~L.}\ \bibnamefont {Gervasio}},\
  }\bibfield  {title} {\bibinfo {title} {Deep learning path-like collective
  variable for enhanced sampling molecular dynamics},\ }\href@noop {}
  {\bibfield  {journal} {\bibinfo  {journal} {J. Chem. Phys.}\ }\textbf
  {\bibinfo {volume} {160}} (\bibinfo {year} {2024})}\BibitemShut {NoStop}%
\bibitem [{\citenamefont {Mondal}\ \emph {et~al.}(2024)\citenamefont {Mondal},
  \citenamefont {Sauer},\ and\ \citenamefont {Heyden}}]{mondal2024exploring}%
  \BibitemOpen
  \bibfield  {author} {\bibinfo {author} {\bibfnamefont {S.}~\bibnamefont
  {Mondal}}, \bibinfo {author} {\bibfnamefont {M.~A.}\ \bibnamefont {Sauer}},\
  and\ \bibinfo {author} {\bibfnamefont {M.}~\bibnamefont {Heyden}},\
  }\bibfield  {title} {\bibinfo {title} {Exploring conformational landscapes
  along anharmonic low-frequency vibrations},\ }\href@noop {} {\bibfield
  {journal} {\bibinfo  {journal} {J. Phys. Chem. B}\ }\textbf {\bibinfo
  {volume} {128}},\ \bibinfo {pages} {7112} (\bibinfo {year}
  {2024})}\BibitemShut {NoStop}%
\bibitem [{\citenamefont {Sauer}\ \emph {et~al.}(2025)\citenamefont {Sauer},
  \citenamefont {Mondal}, \citenamefont {Cano},\ and\ \citenamefont
  {Heyden}}]{sauer2025high}%
  \BibitemOpen
  \bibfield  {author} {\bibinfo {author} {\bibfnamefont {M.~A.}\ \bibnamefont
  {Sauer}}, \bibinfo {author} {\bibfnamefont {S.}~\bibnamefont {Mondal}},
  \bibinfo {author} {\bibfnamefont {M.}~\bibnamefont {Cano}},\ and\ \bibinfo
  {author} {\bibfnamefont {M.}~\bibnamefont {Heyden}},\ }\bibfield  {title}
  {\bibinfo {title} {High-throughput computation of anharmonic low-frequency
  protein vibrations},\ }\href@noop {} {\bibfield  {journal} {\bibinfo
  {journal} {J. Phys. Chem. B}\ }\textbf {\bibinfo {volume} {129}},\ \bibinfo
  {pages} {10739} (\bibinfo {year} {2025})}\BibitemShut {NoStop}%
\bibitem [{\citenamefont {Sauer}\ \emph
  {et~al.}(2026{\natexlab{a}})\citenamefont {Sauer}, \citenamefont {Mondal},
  \citenamefont {Neff}, \citenamefont {Maiti},\ and\ \citenamefont
  {Heyden}}]{sauer2026fast}%
  \BibitemOpen
  \bibfield  {author} {\bibinfo {author} {\bibfnamefont {M.~A.}\ \bibnamefont
  {Sauer}}, \bibinfo {author} {\bibfnamefont {S.}~\bibnamefont {Mondal}},
  \bibinfo {author} {\bibfnamefont {B.}~\bibnamefont {Neff}}, \bibinfo {author}
  {\bibfnamefont {S.}~\bibnamefont {Maiti}},\ and\ \bibinfo {author}
  {\bibfnamefont {M.}~\bibnamefont {Heyden}},\ }\bibfield  {title} {\bibinfo
  {title} {Fast sampling of protein conformational dynamics},\ }\href@noop {}
  {\bibfield  {journal} {\bibinfo  {journal} {Science Advances}\ }\textbf
  {\bibinfo {volume} {12}},\ \bibinfo {pages} {eaea4617} (\bibinfo {year}
  {2026}{\natexlab{a}})}\BibitemShut {NoStop}%
\bibitem [{\citenamefont {Neff}\ and\ \citenamefont
  {Heyden}(2026)}]{neff26protein}%
  \BibitemOpen
  \bibfield  {author} {\bibinfo {author} {\bibfnamefont {B.}~\bibnamefont
  {Neff}}\ and\ \bibinfo {author} {\bibfnamefont {M.}~\bibnamefont {Heyden}},\
  }\href {https://arxiv.org/abs/2601.02699} {\bibinfo {title} {Protein-water
  energy transfer via anharmonic low-frequency vibrations}} (\bibinfo {year}
  {2026}),\ \Eprint {https://arxiv.org/abs/2601.02699} {arXiv:2601.02699
  [cond-mat.soft]} \BibitemShut {NoStop}%
\bibitem [{\citenamefont {McCammon}(1984)}]{mccammon1984protein}%
  \BibitemOpen
  \bibfield  {author} {\bibinfo {author} {\bibfnamefont {J.}~\bibnamefont
  {McCammon}},\ }\bibfield  {title} {\bibinfo {title} {Protein dynamics},\
  }\href@noop {} {\bibfield  {journal} {\bibinfo  {journal} {Rep. Prog. Phys.}\
  }\textbf {\bibinfo {volume} {47}},\ \bibinfo {pages} {1} (\bibinfo {year}
  {1984})}\BibitemShut {NoStop}%
\bibitem [{\citenamefont {Hornak}\ \emph {et~al.}(2006)\citenamefont {Hornak},
  \citenamefont {Okur}, \citenamefont {Rizzo},\ and\ \citenamefont
  {Simmerling}}]{hornak2006hiv}%
  \BibitemOpen
  \bibfield  {author} {\bibinfo {author} {\bibfnamefont {V.}~\bibnamefont
  {Hornak}}, \bibinfo {author} {\bibfnamefont {A.}~\bibnamefont {Okur}},
  \bibinfo {author} {\bibfnamefont {R.~C.}\ \bibnamefont {Rizzo}},\ and\
  \bibinfo {author} {\bibfnamefont {C.}~\bibnamefont {Simmerling}},\ }\bibfield
   {title} {\bibinfo {title} {Hiv-1 protease flaps spontaneously open and
  reclose in molecular dynamics simulations},\ }\href@noop {} {\bibfield
  {journal} {\bibinfo  {journal} {Proc. Natl. Acad. Sci. U.S.A}\ }\textbf
  {\bibinfo {volume} {103}},\ \bibinfo {pages} {915} (\bibinfo {year}
  {2006})}\BibitemShut {NoStop}%
\bibitem [{\citenamefont {Pantsar}(2020)}]{pantsar2020current}%
  \BibitemOpen
  \bibfield  {author} {\bibinfo {author} {\bibfnamefont {T.}~\bibnamefont
  {Pantsar}},\ }\bibfield  {title} {\bibinfo {title} {The current understanding
  of kras protein structure and dynamics},\ }\href@noop {} {\bibfield
  {journal} {\bibinfo  {journal} {Comput. Struct. Biotechnol. J.}\ }\textbf
  {\bibinfo {volume} {18}},\ \bibinfo {pages} {189} (\bibinfo {year}
  {2020})}\BibitemShut {NoStop}%
\bibitem [{\citenamefont {Kim}\ \emph {et~al.}(2021)\citenamefont {Kim},
  \citenamefont {Lee}, \citenamefont {Jeong},\ and\ \citenamefont
  {Jang}}]{kim2021oncogenic}%
  \BibitemOpen
  \bibfield  {author} {\bibinfo {author} {\bibfnamefont {H.~J.}\ \bibnamefont
  {Kim}}, \bibinfo {author} {\bibfnamefont {H.~N.}\ \bibnamefont {Lee}},
  \bibinfo {author} {\bibfnamefont {M.~S.}\ \bibnamefont {Jeong}},\ and\
  \bibinfo {author} {\bibfnamefont {S.~B.}\ \bibnamefont {Jang}},\ }\bibfield
  {title} {\bibinfo {title} {Oncogenic kras: signaling and drug resistance},\
  }\href@noop {} {\bibfield  {journal} {\bibinfo  {journal} {Cancers}\ }\textbf
  {\bibinfo {volume} {13}},\ \bibinfo {pages} {5599} (\bibinfo {year}
  {2021})}\BibitemShut {NoStop}%
\bibitem [{\citenamefont {Jani}\ \emph {et~al.}(2024)\citenamefont {Jani},
  \citenamefont {Sonavane},\ and\ \citenamefont {Joshi}}]{jani2024insight}%
  \BibitemOpen
  \bibfield  {author} {\bibinfo {author} {\bibfnamefont {V.}~\bibnamefont
  {Jani}}, \bibinfo {author} {\bibfnamefont {U.}~\bibnamefont {Sonavane}},\
  and\ \bibinfo {author} {\bibfnamefont {R.}~\bibnamefont {Joshi}},\ }\bibfield
   {title} {\bibinfo {title} {Insight into structural dynamics involved in
  activation mechanism of full length kras wild type and p-loop mutants},\
  }\href@noop {} {\bibfield  {journal} {\bibinfo  {journal} {Heliyon}\ }\textbf
  {\bibinfo {volume} {10}} (\bibinfo {year} {2024})}\BibitemShut {NoStop}%
\bibitem [{\citenamefont {Bery}\ \emph {et~al.}(2019)\citenamefont {Bery},
  \citenamefont {Legg}, \citenamefont {Debreczeni}, \citenamefont {Breed},
  \citenamefont {Embrey}, \citenamefont {Stubbs}, \citenamefont
  {Kolasinska-Zwierz}, \citenamefont {Barrett}, \citenamefont {Marwood},
  \citenamefont {Watson} \emph {et~al.}}]{bery2019kras}%
  \BibitemOpen
  \bibfield  {author} {\bibinfo {author} {\bibfnamefont {N.}~\bibnamefont
  {Bery}}, \bibinfo {author} {\bibfnamefont {S.}~\bibnamefont {Legg}}, \bibinfo
  {author} {\bibfnamefont {J.}~\bibnamefont {Debreczeni}}, \bibinfo {author}
  {\bibfnamefont {J.}~\bibnamefont {Breed}}, \bibinfo {author} {\bibfnamefont
  {K.}~\bibnamefont {Embrey}}, \bibinfo {author} {\bibfnamefont
  {C.}~\bibnamefont {Stubbs}}, \bibinfo {author} {\bibfnamefont
  {P.}~\bibnamefont {Kolasinska-Zwierz}}, \bibinfo {author} {\bibfnamefont
  {N.}~\bibnamefont {Barrett}}, \bibinfo {author} {\bibfnamefont
  {R.}~\bibnamefont {Marwood}}, \bibinfo {author} {\bibfnamefont
  {J.}~\bibnamefont {Watson}}, \emph {et~al.},\ }\bibfield  {title} {\bibinfo
  {title} {Kras-specific inhibition using a darpin binding to a site in the
  allosteric lobe},\ }\href@noop {} {\bibfield  {journal} {\bibinfo  {journal}
  {Nat. Commun.}\ }\textbf {\bibinfo {volume} {10}},\ \bibinfo {pages} {2607}
  (\bibinfo {year} {2019})}\BibitemShut {NoStop}%
\bibitem [{\citenamefont {Costa}\ \emph {et~al.}(2023)\citenamefont {Costa},
  \citenamefont {Batista}, \citenamefont {Gomes}, \citenamefont {Bastos},
  \citenamefont {Louet}, \citenamefont {Floquet}, \citenamefont {Bisch},\ and\
  \citenamefont {Perahia}}]{costa2023}%
  \BibitemOpen
  \bibfield  {author} {\bibinfo {author} {\bibfnamefont {M.~G.}\ \bibnamefont
  {Costa}}, \bibinfo {author} {\bibfnamefont {P.~R.}\ \bibnamefont {Batista}},
  \bibinfo {author} {\bibfnamefont {A.}~\bibnamefont {Gomes}}, \bibinfo
  {author} {\bibfnamefont {L.~S.}\ \bibnamefont {Bastos}}, \bibinfo {author}
  {\bibfnamefont {M.}~\bibnamefont {Louet}}, \bibinfo {author} {\bibfnamefont
  {N.}~\bibnamefont {Floquet}}, \bibinfo {author} {\bibfnamefont {P.~M.}\
  \bibnamefont {Bisch}},\ and\ \bibinfo {author} {\bibfnamefont
  {D.}~\bibnamefont {Perahia}},\ }\bibfield  {title} {\bibinfo {title}
  {Mdexciter: Enhanced sampling molecular dynamics by excited normal modes or
  principal components obtained from experiments},\ }\href@noop {} {\bibfield
  {journal} {\bibinfo  {journal} {J. Chem. Theory Comput.}\ }\textbf {\bibinfo
  {volume} {19}},\ \bibinfo {pages} {412} (\bibinfo {year} {2023})}\BibitemShut
  {NoStop}%
\bibitem [{\citenamefont {Huang}\ \emph {et~al.}(2018)\citenamefont {Huang},
  \citenamefont {McCammon},\ and\ \citenamefont {Miao}}]{huang2018replica}%
  \BibitemOpen
  \bibfield  {author} {\bibinfo {author} {\bibfnamefont {Y.~M.}\ \bibnamefont
  {Huang}}, \bibinfo {author} {\bibfnamefont {J.~A.}\ \bibnamefont
  {McCammon}},\ and\ \bibinfo {author} {\bibfnamefont {Y.}~\bibnamefont
  {Miao}},\ }\bibfield  {title} {\bibinfo {title} {Replica exchange gaussian
  accelerated molecular dynamics: Improved enhanced sampling and free energy
  calculation},\ }\href@noop {} {\bibfield  {journal} {\bibinfo  {journal} {J.
  Chem. Theory Comput.}\ }\textbf {\bibinfo {volume} {14}},\ \bibinfo {pages}
  {1853} (\bibinfo {year} {2018})}\BibitemShut {NoStop}%
\bibitem [{\citenamefont {Ren}\ \emph {et~al.}(2021)\citenamefont {Ren},
  \citenamefont {Dokainish}, \citenamefont {Shinobu}, \citenamefont {Oshima},\
  and\ \citenamefont {Sugita}}]{ren2021unraveling}%
  \BibitemOpen
  \bibfield  {author} {\bibinfo {author} {\bibfnamefont {W.}~\bibnamefont
  {Ren}}, \bibinfo {author} {\bibfnamefont {H.~M.}\ \bibnamefont {Dokainish}},
  \bibinfo {author} {\bibfnamefont {A.}~\bibnamefont {Shinobu}}, \bibinfo
  {author} {\bibfnamefont {H.}~\bibnamefont {Oshima}},\ and\ \bibinfo {author}
  {\bibfnamefont {Y.}~\bibnamefont {Sugita}},\ }\bibfield  {title} {\bibinfo
  {title} {Unraveling the coupling between conformational changes and ligand
  binding in ribose binding protein using multiscale molecular dynamics and
  free-energy calculations},\ }\href@noop {} {\bibfield  {journal} {\bibinfo
  {journal} {J. Phys. Chem. B}\ }\textbf {\bibinfo {volume} {125}},\ \bibinfo
  {pages} {2898} (\bibinfo {year} {2021})}\BibitemShut {NoStop}%
\bibitem [{\citenamefont {Benabderrahmane}\ \emph {et~al.}(2020)\citenamefont
  {Benabderrahmane}, \citenamefont {Bureau}, \citenamefont {Voisin-Chiret},\
  and\ \citenamefont {Sopkova-de
  Oliveira~Santos}}]{benabderrahmane2020insights}%
  \BibitemOpen
  \bibfield  {author} {\bibinfo {author} {\bibfnamefont {M.}~\bibnamefont
  {Benabderrahmane}}, \bibinfo {author} {\bibfnamefont {R.}~\bibnamefont
  {Bureau}}, \bibinfo {author} {\bibfnamefont {A.~S.}\ \bibnamefont
  {Voisin-Chiret}},\ and\ \bibinfo {author} {\bibfnamefont {J.}~\bibnamefont
  {Sopkova-de Oliveira~Santos}},\ }\bibfield  {title} {\bibinfo {title}
  {Insights into mcl-1 conformational states and allosteric inhibition
  mechanism from molecular dynamics simulations, enhanced sampling, and pocket
  crosstalk analysis},\ }\href@noop {} {\bibfield  {journal} {\bibinfo
  {journal} {J. Chem. Inf. Model.}\ }\textbf {\bibinfo {volume} {60}},\
  \bibinfo {pages} {3172} (\bibinfo {year} {2020})}\BibitemShut {NoStop}%
\bibitem [{\citenamefont {Brooks}\ and\ \citenamefont
  {Karplus}(1983)}]{brooks1983harmonic}%
  \BibitemOpen
  \bibfield  {author} {\bibinfo {author} {\bibfnamefont {B.}~\bibnamefont
  {Brooks}}\ and\ \bibinfo {author} {\bibfnamefont {M.}~\bibnamefont
  {Karplus}},\ }\bibfield  {title} {\bibinfo {title} {Harmonic dynamics of
  proteins: normal modes and fluctuations in bovine pancreatic trypsin
  inhibitor.},\ }\href@noop {} {\bibfield  {journal} {\bibinfo  {journal}
  {Proc. Natl. Acad. Sci. U.S.A}\ }\textbf {\bibinfo {volume} {80}},\ \bibinfo
  {pages} {6571} (\bibinfo {year} {1983})}\BibitemShut {NoStop}%
\bibitem [{\citenamefont {Ribeiro}\ \emph {et~al.}(2018)\citenamefont
  {Ribeiro}, \citenamefont {Bravo}, \citenamefont {Wang},\ and\ \citenamefont
  {Tiwary}}]{ribeiro2018reweighted}%
  \BibitemOpen
  \bibfield  {author} {\bibinfo {author} {\bibfnamefont {J.~M.~L.}\
  \bibnamefont {Ribeiro}}, \bibinfo {author} {\bibfnamefont {P.}~\bibnamefont
  {Bravo}}, \bibinfo {author} {\bibfnamefont {Y.}~\bibnamefont {Wang}},\ and\
  \bibinfo {author} {\bibfnamefont {P.}~\bibnamefont {Tiwary}},\ }\bibfield
  {title} {\bibinfo {title} {Reweighted autoencoded variational bayes for
  enhanced sampling (rave)},\ }\href@noop {} {\bibfield  {journal} {\bibinfo
  {journal} {J. Chem. Phys.}\ }\textbf {\bibinfo {volume} {149}} (\bibinfo
  {year} {2018})}\BibitemShut {NoStop}%
\bibitem [{\citenamefont {Wang}\ \emph {et~al.}(2019)\citenamefont {Wang},
  \citenamefont {Ribeiro},\ and\ \citenamefont {Tiwary}}]{wang2019past}%
  \BibitemOpen
  \bibfield  {author} {\bibinfo {author} {\bibfnamefont {Y.}~\bibnamefont
  {Wang}}, \bibinfo {author} {\bibfnamefont {J.~M.~L.}\ \bibnamefont
  {Ribeiro}},\ and\ \bibinfo {author} {\bibfnamefont {P.}~\bibnamefont
  {Tiwary}},\ }\bibfield  {title} {\bibinfo {title} {Past--future information
  bottleneck for sampling molecular reaction coordinate simultaneously with
  thermodynamics and kinetics},\ }\href@noop {} {\bibfield  {journal} {\bibinfo
   {journal} {Nat. Commun.}\ }\textbf {\bibinfo {volume} {10}},\ \bibinfo
  {pages} {3573} (\bibinfo {year} {2019})}\BibitemShut {NoStop}%
\bibitem [{\citenamefont {Istomin}\ and\ \citenamefont
  {Piccini}(2024)}]{istomin2024festa}%
  \BibitemOpen
  \bibfield  {author} {\bibinfo {author} {\bibfnamefont {V.}~\bibnamefont
  {Istomin}}\ and\ \bibinfo {author} {\bibfnamefont {G.}~\bibnamefont
  {Piccini}},\ }\bibfield  {title} {\bibinfo {title} {Festa: A polygon-based
  approach for extracting relevant structures from free energy surfaces
  obtained in molecular simulations},\ }\href@noop {} {\bibfield  {journal}
  {\bibinfo  {journal} {J. Chem. Inf. Model.}\ }\textbf {\bibinfo {volume}
  {65}},\ \bibinfo {pages} {1} (\bibinfo {year} {2024})}\BibitemShut {NoStop}%
\bibitem [{\citenamefont {Chatterjee}\ and\ \citenamefont
  {Ray}(2025)}]{chatterjee2025acceleration}%
  \BibitemOpen
  \bibfield  {author} {\bibinfo {author} {\bibfnamefont {S.}~\bibnamefont
  {Chatterjee}}\ and\ \bibinfo {author} {\bibfnamefont {D.}~\bibnamefont
  {Ray}},\ }\bibfield  {title} {\bibinfo {title} {Acceleration with
  interpretability: A surrogate model-based collective variable for enhanced
  sampling},\ }\href@noop {} {\bibfield  {journal} {\bibinfo  {journal} {J.
  Chem. Theory Comput.}\ }\textbf {\bibinfo {volume} {21}},\ \bibinfo {pages}
  {1561} (\bibinfo {year} {2025})}\BibitemShut {NoStop}%
\bibitem [{\citenamefont {De~Simone}\ \emph {et~al.}(2013)\citenamefont
  {De~Simone}, \citenamefont {Montalvao}, \citenamefont {Dobson},\ and\
  \citenamefont {Vendruscolo}}]{desimone2013characterization}%
  \BibitemOpen
  \bibfield  {author} {\bibinfo {author} {\bibfnamefont {A.}~\bibnamefont
  {De~Simone}}, \bibinfo {author} {\bibfnamefont {R.~W.}\ \bibnamefont
  {Montalvao}}, \bibinfo {author} {\bibfnamefont {C.~M.}\ \bibnamefont
  {Dobson}},\ and\ \bibinfo {author} {\bibfnamefont {M.}~\bibnamefont
  {Vendruscolo}},\ }\bibfield  {title} {\bibinfo {title} {Characterization of
  the interdomain motions in hen lysozyme using residual dipolar couplings as
  replica-averaged structural restraints in molecular dynamics simulations},\
  }\href@noop {} {\bibfield  {journal} {\bibinfo  {journal} {Biochemistry}\
  }\textbf {\bibinfo {volume} {52}},\ \bibinfo {pages} {6480} (\bibinfo {year}
  {2013})}\BibitemShut {NoStop}%
\bibitem [{\citenamefont {Chen}\ \emph {et~al.}(2021)\citenamefont {Chen},
  \citenamefont {Zhang}, \citenamefont {Wang}, \citenamefont {Pang},
  \citenamefont {Zhang},\ and\ \citenamefont {Liu}}]{chen2021mutation}%
  \BibitemOpen
  \bibfield  {author} {\bibinfo {author} {\bibfnamefont {J.}~\bibnamefont
  {Chen}}, \bibinfo {author} {\bibfnamefont {S.}~\bibnamefont {Zhang}},
  \bibinfo {author} {\bibfnamefont {W.}~\bibnamefont {Wang}}, \bibinfo {author}
  {\bibfnamefont {L.}~\bibnamefont {Pang}}, \bibinfo {author} {\bibfnamefont
  {Q.}~\bibnamefont {Zhang}},\ and\ \bibinfo {author} {\bibfnamefont
  {X.}~\bibnamefont {Liu}},\ }\bibfield  {title} {\bibinfo {title}
  {Mutation-induced impacts on the switch transformations of the gdp-and
  gtp-bound k-ras: insights from multiple replica gaussian accelerated
  molecular dynamics and free energy analysis},\ }\href@noop {} {\bibfield
  {journal} {\bibinfo  {journal} {J. Chem. Inf. Model.}\ }\textbf {\bibinfo
  {volume} {61}},\ \bibinfo {pages} {1954} (\bibinfo {year}
  {2021})}\BibitemShut {NoStop}%
\bibitem [{\citenamefont {Barducci}\ \emph {et~al.}(2008)\citenamefont
  {Barducci}, \citenamefont {Bussi},\ and\ \citenamefont
  {Parrinello}}]{barducci2008well}%
  \BibitemOpen
  \bibfield  {author} {\bibinfo {author} {\bibfnamefont {A.}~\bibnamefont
  {Barducci}}, \bibinfo {author} {\bibfnamefont {G.}~\bibnamefont {Bussi}},\
  and\ \bibinfo {author} {\bibfnamefont {M.}~\bibnamefont {Parrinello}},\
  }\bibfield  {title} {\bibinfo {title} {Well-tempered metadynamics: a smoothly
  converging and tunable free-energy method},\ }\href@noop {} {\bibfield
  {journal} {\bibinfo  {journal} {Phys. Rev. Lett.}\ }\textbf {\bibinfo
  {volume} {100}},\ \bibinfo {pages} {020603} (\bibinfo {year}
  {2008})}\BibitemShut {NoStop}%
\bibitem [{\citenamefont {Sauer}\ \emph
  {et~al.}(2026{\natexlab{b}})\citenamefont {Sauer}, \citenamefont {Mondal},\
  and\ \citenamefont {Heyden}}]{FRESEANmetadynamics}%
  \BibitemOpen
  \bibfield  {author} {\bibinfo {author} {\bibfnamefont {M.~A.}\ \bibnamefont
  {Sauer}}, \bibinfo {author} {\bibfnamefont {S.}~\bibnamefont {Mondal}},\ and\
  \bibinfo {author} {\bibfnamefont {M.}~\bibnamefont {Heyden}},\ }\bibfield
  {title} {\bibinfo {title} {Fresean metadynamics with weighted {DCCM}
  analysis}\ }\href {https://doi.org/DOI: 10.5281/zenodo.20076452} {DOI:
  10.5281/zenodo.20076452} (\bibinfo {year} {2026}{\natexlab{b}})\BibitemShut
  {NoStop}%
\end{thebibliography}
%

\clearpage

\pagebreak
\widetext
\newpage
\begin{center}
\textbf{\large Supporting Information: From Enhanced Sampling to Human-Readable Representations of Protein Dynamics}
\end{center}
\setcounter{equation}{0}
\setcounter{figure}{0}
\setcounter{table}{0}
\setcounter{page}{1}
\makeatletter
\renewcommand{\theequation}{S\arabic{equation}}
\renewcommand{\thefigure}{S\arabic{figure}}
\renewcommand{\thetable}{S\arabic{table}}

\begin{figure}[ht!]
\begin{center}
\includegraphics[clip=true,width=0.75\linewidth]{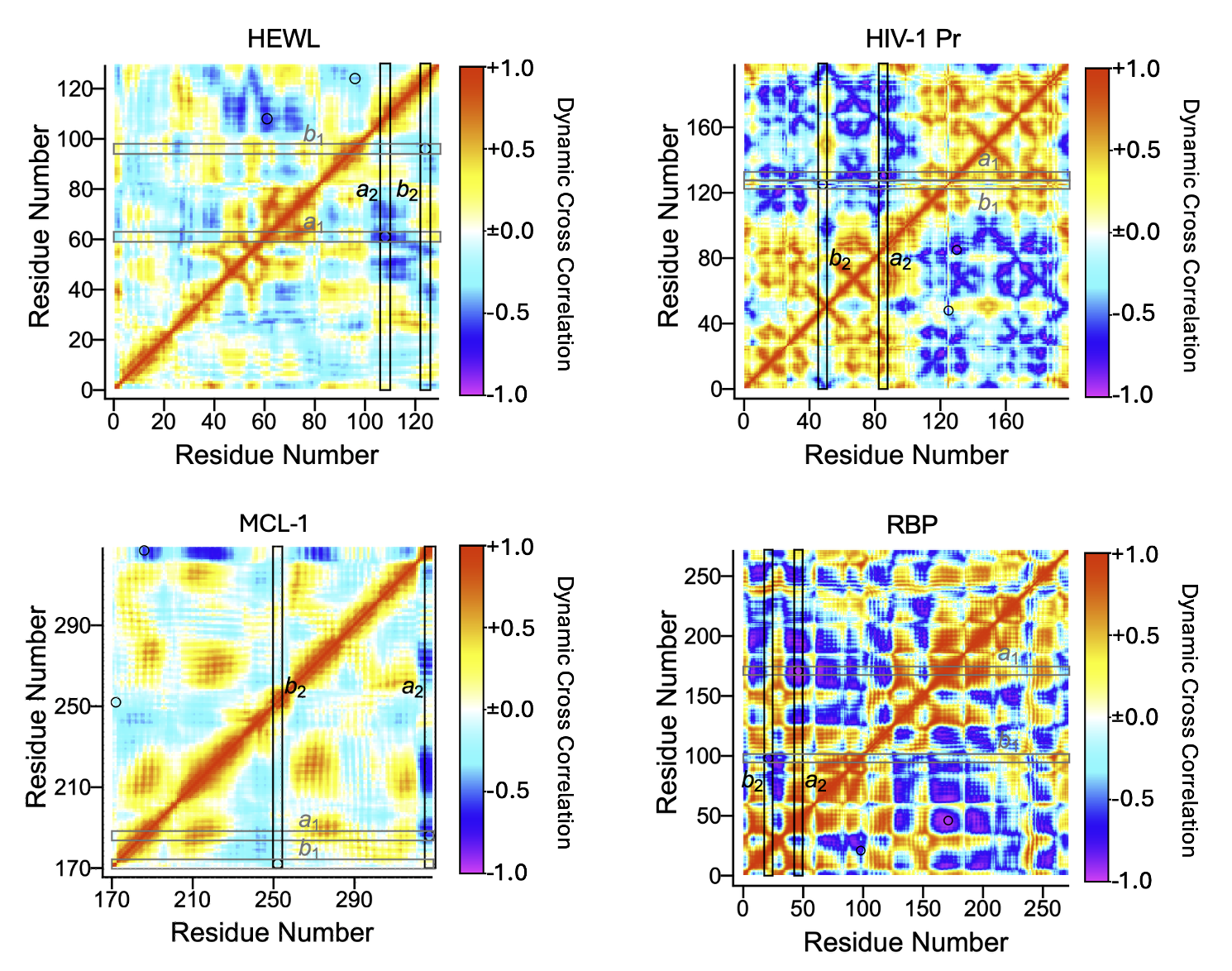}
\caption{
Dynamic cross-correlation matrices (DCCMs) computed as weighted averages from enhanced sampling simulations for HEWL, HIV-1 protease, MCL-1, and ribose-binding protein (RBP).
Each DCCM is constructed from biased trajectories using the weighting procedure described in the Methods and Theory sections.
Matrix elements range from -1 to +1, indicating anti-correlated and correlated motions, respectively. 
These results are analogous to the KRAS DCCM shown in Figure 1 of the main text.
}
\label{f:DCCM-SI}
\end{center}
\end{figure}

\begin{figure}[ht!]
\begin{center}
\includegraphics[clip=true,width=0.75\linewidth]{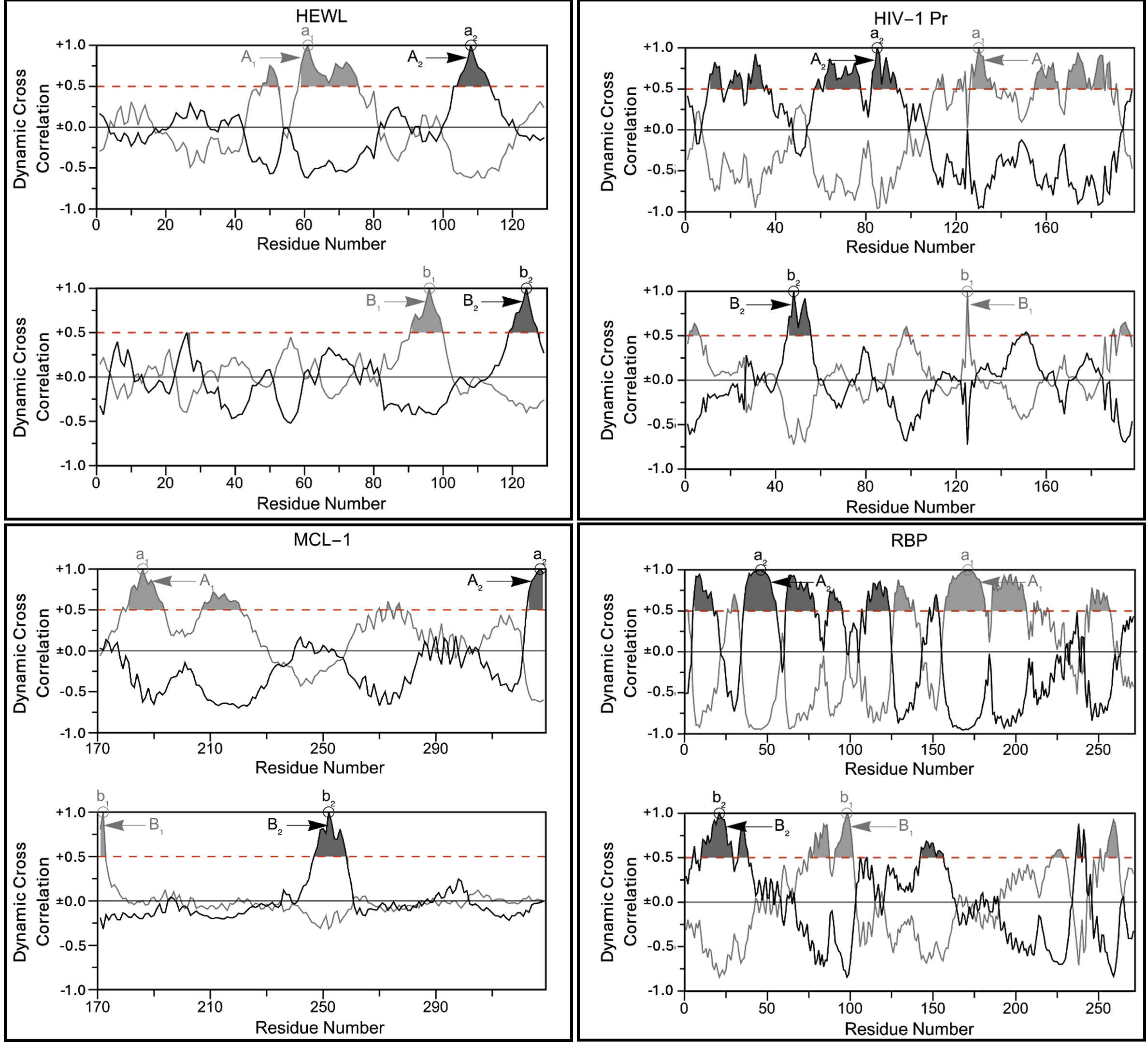}
\caption{
Numerical values in selected rows and columns of the DCCMs for HEWL, HIV-1 protease, MCL-1, and ribose-binding protein (RBP), corresponding to the residues identified by the selection algorithm described in the Theory section.
For each system, rows/columns associated with residues $a_1$, $a_2$ and $b_1$, $b_2$ are shown (light and dark gray, respectively).
The plotted values quantify correlations of all residues with the selected residues, with self-correlations equal to 1 by definition.
A threshold criterion (dashed line) is used to identify residues forming the collectively moving domains $A_1$, $A_2$, $B_1$, and $B_2$, highlighted by shaded regions.
These plots are analogous to those shown for KRAS in Figure 2 of the main text.
}
\label{f:DCCM-rows-SI}
\end{center}
\end{figure}

\begin{figure}[ht!]
\begin{center}
\includegraphics[clip=true,width=0.75\linewidth]{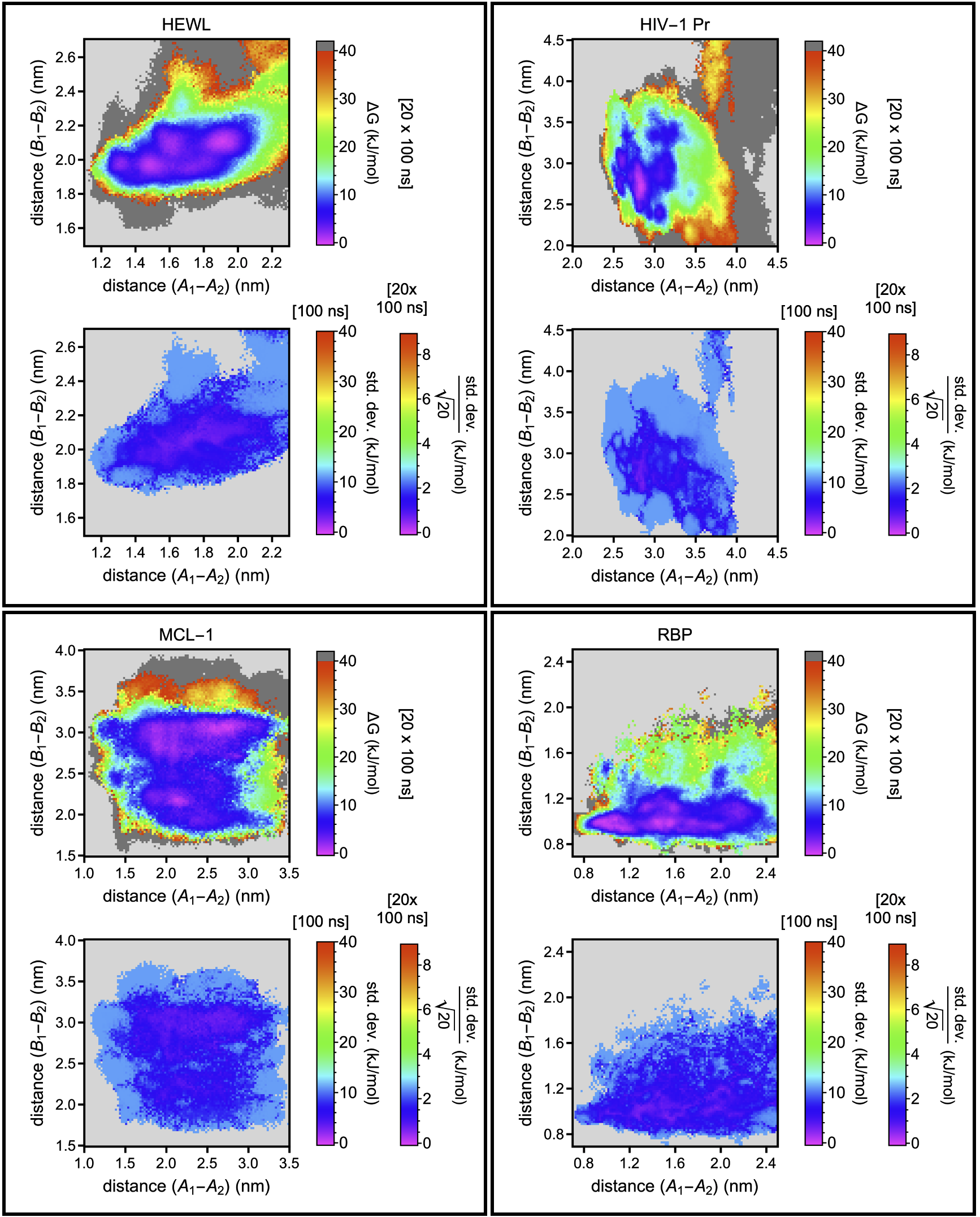}
\caption{
Free energy surfaces for HEWL, HIV-1 protease, MCL-1, and ribose-binding protein (RBP) as functions of the domain-domain distances $d_A=\left|A_1-A_2\right|$ and $d_B=\left|B_1-B_2\right|$, obtained after unbiasing enhanced sampling trajectories.
For each system, free energy surfaces are constructed from weighted histograms averaged over multiple independent metadynamics simulations, as described in the Methods section.
The resulting surfaces provide human-readable representations of the conformational ensembles and are directly comparable to the KRAS results shown in Figure 5 of the main text.
}
\label{f:FES-SI}
\end{center}
\end{figure}

\end{document}